\documentclass{article}

\usepackage{PRIMEarxiv}

\usepackage[utf8]{inputenc} 
\usepackage[T1]{fontenc}    
\usepackage{hyperref}       
\usepackage{url}            
\usepackage{booktabs}       
\usepackage{amsfonts}       
\usepackage{nicefrac}       
\usepackage{microtype}      
\usepackage{lipsum}
\usepackage{fancyhdr}       
\usepackage{amsmath}
\usepackage{graphicx}       
\graphicspath{{media/}}     
\usepackage{rotating}
\usepackage{graphicx}
\usepackage{xcolor}
\usepackage{lscape}
\usepackage{multirow}
\usepackage{float}
\usepackage{tabularray}
\usepackage{adjustbox}
\usepackage{tabularx,ragged2e,booktabs}
\usepackage{tabulary,booktabs}
\usepackage{tabularray}
\title{\textbf{GRAPHITE}: Graph-Based Interpretable Tissue Examination for Enhanced Explainability in Breast Cancer Histopathology
}

\author{
  Raktim Kumar Mondol \\
    School of Computer Science and Engineering \\
  University of New South Wales  \\
  Sydney, NSW 2052, Australia
   \And
  Ewan K.~A.~Millar\\
  St. George and Sutherland Clinical School\\
  University of New South Wales\\
  NSW 2217, Sydney, Australia\\
  \\
  Department of Anatomical Pathology\\
  NSW Health Pathology, St. George Hospital\\
  Kogarah, NSW 2217, Australia
  \AND
  Peter H.~Graham, Lois Browne \\
  St. George and Sutherland Clinical School\\
  University of New South Wales\\
  NSW 2217, Sydney, Australia\\
  \\
  Cancer Care Centre, St George Hospital \\
  Kogarah, NSW 2217, Australia 
  \AND
  Arcot Sowmya, Erik Meijering \\
  School of Computer Science and Engineering \\
  University of New South Wales  \\
  Sydney, NSW 2052, Australia\\
  \texttt{\{a.sowmya, erik.meijering\}@unsw.edu.au}
}

\begin{document}
\maketitle
\begin{abstract}
\section*{Abstract}

Explainable AI (XAI) in medical histopathology is essential for enhancing the interpretability and clinical trustworthiness of deep learning models in cancer diagnosis. However, the black-box nature of these models often limits their clinical adoption. We introduce GRAPHITE (Graph-based Interpretable Tissue Examination), a post-hoc explainable framework designed for breast cancer tissue microarray (TMA) analysis. GRAPHITE employs a multiscale approach, extracting patches at various magnification levels, constructing an hierarchical graph, and utilising graph attention networks (GAT) with scalewise attention (SAN) to capture scale-dependent features. We trained the model on 140 tumour TMA cores and four benign whole slide images from which 140 benign samples were created, and tested it on 53 pathologist-annotated TMA samples. GRAPHITE outperformed traditional XAI methods, achieving a mean average precision (mAP) of 0.56, an area under the receiver operating characteristic curve (AUROC) of 0.94, and a threshold robustness (ThR) of 0.70, indicating that the model maintains high performance across a wide range of thresholds. In clinical utility, GRAPHITE achieved the highest area under the decision curve (AUDC) of 4.17e+5, indicating reliable decision support across thresholds. These results highlight GRAPHITE’s potential as a clinically valuable tool in computational pathology, providing interpretable visualisations that align with the pathologists' diagnostic reasoning and support precision medicine.

\end{abstract}

\section{Introduction}
Digital pathology has revolutionised diagnostic medicine by enabling the digitisation of histopathology slides, facilitating high-throughput analysis \cite{10.1002/gcc.23192,10.1007/s00428-022-03327-2}. The availability of large-scale image datasets and advances in computational power have catalysed the integration of artificial intelligence (AI) with cancer diagnosis \cite{10.1080/17434440.2019.1610387, 10.3389/fonc.2020.594580}. Deep learning models, in particular, have shown exceptional promise in analysing complex tissue structures and offer automated predictions with high accuracy \cite{10.1038/s41591-018-0177-5, 10.1002/cam4.7252, 10.1038/s41598-021-84510-4, 10.3390/s20164373}. Cancer diagnosis, especially for breast cancer, has benefitted significantly from these technologies, which provide robust decision support in areas such as tumour detection, grading and subtype classification \cite{10.3233/cbm-230251,10.1038/s41598-022-19112-9,10.1038/s41523-022-00478-y,10.1109/tmi.2017.2758580, 10.1109/access.2020.3029881}. These applications of AI in digital pathology not only streamline workflows but also help mitigate challenges posed by the increasing volume of pathology cases, particularly in oncology.

Despite advances in AI for digital pathology, challenges remain in the interpretability of deep learning models \cite{10.1038/s43856-021-00013-3}. These models often operate as ``black boxes,'' making it difficult for pathologists to understand the rationale behind specific predictions. This lack of transparency can hinder the clinical adoption of AI tools, as pathologists may be reluctant to trust a system that does not provide clear explanations for its decisions \cite{10.1155/2023/4637678, 10.1186/s12911-021-01542-6, 10.1186/s12910-022-00842-4}. In addition, the interpretability of AI models is crucial in cancer diagnosis, where erroneous interpretations have significant consequences for patient care. The challenge of interpretability is compounded by the complexity of histopathology images, which contain intricate patterns that may not be easily discernible even to experienced pathologists. Specifically, in breast cancer tissue microarrays (TMAs), tissue heterogeneity further complicates interpretation, requiring models to not only make accurate predictions but also provide meaningful explanations.

Explainable AI (XAI) has emerged as a key area of research addressing these challenges by enhancing the transparency of deep learning models. XAI methods provide visual or textual explanations that align with clinical reasoning, thereby building trust among pathologists and clinicians \cite{10.1038/s43856-023-00276-y, 10.1038/s41698-023-00472-y, 10.1038/s42256-021-00303-4, 10.1016/j.ccell.2022.07.004, 10.1111/coin.12660}. This alignment is essential for fostering confidence in AI-assisted diagnoses and ensuring that pathologists can effectively integrate these tools into their workflows. In cancer diagnosis, this translates to models generating visual explanations that highlight diagnostically relevant regions within tissue samples, allowing pathologists to validate the AI's findings. Furthermore, XAI can facilitate the identification of potential biases in AI models, allowing the refinement of algorithms to ensure equitable and accurate outcomes across diverse patient populations \cite{10.1155/2022/8167821}. Consequently, the development of post-hoc explainability methods tailored to histopathology is crucial to bridge the gap between model performance and clinical usability.

Current limitations in histopathology interpretation include interobserver variability, subjective assessments and the potential for diagnostic errors \cite{10.2340/00015555-1517,10.3906/sag-2006-257}. Even experienced pathologists may arrive at different conclusions when interpreting the same histopathology slides, leading to inconsistencies in diagnoses. This variability can be attributed to factors such as differences in training and experience, and personal biases. Additionally, the subjective nature of histopathology assessments may result in diagnostic errors, particularly in cases where tumour differentiation is subtle. The integration of AI and digital pathology has the potential to address these limitations by providing objective, data-driven analyses that can provide more consistent results. Approaches such as Grad-CAM \cite{10.1007/s11263-019-01228-7}, which have been widely used in medical imaging, often provide coarse explanations that do not capture the intricate relationships between tissue components. Similarly, attention-based methods face challenges in capturing broader contextual information, potentially leading to less effective feature representation, which is critical for understanding cancer tissue \cite{Di_2024, Hekal_2024}. These limitations highlight the need for novel explainable approaches that can adapt to the hierarchical and multiscale nature of histopathology data.

In this work, we introduce GRAPHITE (Graph-based Interpretable Tissue Examination), a novel post-hoc explainability framework specifically designed for breast cancer TMA analysis. Unlike dynamic graph methods \cite{10.1109/access.2024.3401460}, GRAPHITE employs a static multiscale hierarchical graph representation, capturing tissue-level dependencies across fixed magnification levels. This ensures the preservation of both global and local relationships while maintaining computational efficiency. Additionally, GRAPHITE integrates gradient-based heat maps with graph-based attention to offer more interpretable visualisations, highlighting critical tissue regions essential for diagnosis. We also introduce two novel evaluation metrics: threshold stability (ThS), which measures the consistency of model performance across different thresholds by analysing F1 score variations, and threshold robustness (ThR), which quantifies the range of thresholds where the model maintains 95\% of its peak F1 performance. The key contributions of this paper are as follows:

\begin{enumerate}
    \item \textbf{Scale-Aware Graph Learning:} 
    GRAPHITE constructs hierarchical graphs from multiscale tissue patches, processes them through graph attention networks (GAT) and a scalewise attention network (SAN) to generate level-specific attention maps.
    
    \item \textbf{Integration of Multiple Saliency Maps:} 
   GRAPHITE generates comprehensive visualisations through a two-stage fusion process: first combining level-specific attention maps from the SAN into a combined representation, then adaptively fusing it with multiple instance learning (MIL) attention and gradient-based maps using confidence scores to create clinically interpretable saliency maps.
    
    \item \textbf{Comprehensive Evaluation of Visualisation Performance:} 
    \\GRAPHITE is rigorously evaluated using both standard performance metrics and the newly proposed metrics (ThS and ThR), demonstrating superior results compared to existing attention and gradient based methods across all evaluation criteria.
\end{enumerate}

Through these contributions, GRAPHITE addresses the interpretability gap in histopathology and sets a new benchmark for explainability in breast cancer diagnosis. The proposed framework not only improves visualisation quality but also has the potential to enhance clinical trust and adoption, facilitating more reliable computer-aided diagnosis in pathology.

\section{Literature Review}
The adoption of deep learning models in medical imaging has increased the demand for XAI techniques to enhance transparency in the decision-making process and improve trust and effective diagnostic support \cite{10.1016/j.heliyon.2023.e16110, 10.1080/07853890.2023.2233541}. Here we briefly review model-specific XAI techniques, including gradient-based methods and attention mechanisms, analyzing their capabilities and limitations in clinical applications.

Gradient-based visualization techniques have emerged as foundational approaches for interpreting convolutional neural networks (CNNs) in medical imaging. The Gradient-weighted Class Activation Mapping (Grad-CAM) \cite{10.1007/s11263-019-01228-7} generates visual explanations by utilizing gradients flowing into the final convolutional layer. While widely adopted across medical specialties including dermatology, ophthalmology, radiology, and histopathology \cite{10.3390/diagnostics14070753, 10.1109/access.2024.3359698, 10.1167/tvst.9.2.20, 10.3390/diagnostics12092084, 10.1038/s42256-022-00536-x}, Grad-CAM's reliance on the last convolutional layer can limit its precision in high-resolution image analysis \cite{10.1007/s11263-019-01228-7}. The technique also struggles with detecting multiple instances of the same class and providing accurate object localization. To address these limitations, Grad-CAM++ introduced pixel-wise weighting of gradients to improve localization accuracy and handle multiple class occurrences \cite{10.1109/WACV.2018.00097}. Its emphasis on positive gradient contributions enhances localization accuracy. However, the method's increased computational complexity can pose challenges for real-time clinical applications. Despite its improvements, Grad-CAM++ may not always capture the most relevant features, particularly in histopathology where tissue structure granularity is crucial for accurate diagnosis~\cite{10.1109/WACV.2018.00097, van_der_Velden_2022}.

FullGrad offered a more comprehensive approach by considering both forward and backward network passes~\cite{fullgrad}. It integrates input gradients with neuron-specific sensitivities to generate comprehensive saliency maps. However, its reliance on input attribution combined with neuron-level aggregation may increase computational overhead. Other CAM variants have made significant contributions to reduce this computational overhead. For example, EigenCAM \cite{10.1109/IJCNN48605.2020.9206626} employs eigenvalue decomposition of feature maps to generate explanations with improved computational efficiency. While this approach reduces computational overhead, it may not capture fine-grained details crucial for medical diagnosis. Finally, Ablation-CAM circumvents the gradient dependence by evaluating feature map importance through ablation analysis, providing class-discriminative and gradient-free explanations\cite{10.1109/WACV45572.2020.9093360}. It demonstrates superior performance in scenarios where gradient-based methods fail, such as high-confidence predictions and highlighting multiple instances. However, Ablation-CAM may require additional computational resources due to its ablation-driven approach.

Attention mechanisms provide a powerful approach to improving interpretability in medical imaging by enabling models to focus on relevant image regions, thereby highlighting essential features and ignoring irrelevant background \cite{attentionviz, 10.53941/ijndi0201006, 10.1155/2023/7464628}. This is particularly beneficial in applications requiring fine discrimination of subtle tissue differences, such as tumor detection and tissue segmentation. In radiology, attention maps can identify pertinent areas in X-ray or MRI scans, aiding radiologists in their assessments \cite{10.3389/fmed.2023.1180773}. In optical coherence tomography (OCT) image analysis, attention mechanisms have enhanced retinal image segmentation, offering clearer insights into pathological conditions \cite{10.1097/apo.0000000000000405, 10.1038/s42003-021-01697-y, 10.1371/journal.pone.0296175, 10.1371/journal.pone.0287301}. By visualizing biases within AI models, attention-based methods address potential imbalances in training datasets, promoting fairness in healthcare applications \cite{10.1016/j.heliyon.2023.e16110, 10.3389/fnagi.2023.1238065}. Additionally, multiscale attention mechanisms improve interpretability by capturing both local and global contextual information, enhancing the model’s understanding of complex anatomical structures \cite{10.1109/access.2021.3135637}. However, attention mechanisms can introduce ambiguity, as highlighted regions may not always correspond to clinically relevant features, necessitating careful validation to ensure alignment with clinical expectations \cite{10.32604/cmc.2021.017481, 10.1136/bmjophth-2023-001411}. Furthermore, attention mechanisms, especially self-attention, are computationally demanding, which can pose barriers to their clinical adoption, particularly in resource-limited settings \cite{10.1088/1361-6560/ace6f1}.

Beyond gradient-based methods, model-agnostic techniques like LIME (Local Interpretable Model-agnostic Explanations) \cite{lime} and SHAP (SHapley Additive exPlanations) \cite{shap} offer alternative approaches to interpretability. LIME explains individual predictions by learning a simpler, interpretable model locally around the prediction instance. SHAP uses a game-theoretic approach to assign contribution values to each feature for a prediction. These methods have been applied in medical imaging, including breast cancer histopathology \cite{Murugan2024}, offering flexibility as they can be applied to any underlying classifier.

Evaluating XAI methods in medical imaging poses unique challenges, especially for complex datasets such as whole-slide images in histopathology. The lack of standardized evaluation frameworks continues to hinder the comparison and validation of different XAI approaches, creating a significant barrier to their broader adoption  in clinical practice~\cite{10.1097/ICU.0000000000000983, 10.3390/diagnostics12020237}. Current evaluation approaches often depend on qualitative assessments by domain experts and computational metrics like Intersection over Union (IoU) and pointing game to measure localization accuracy by comparing explanations to ground truth annotations \cite{10.1007/s11263-019-01228-7}. However, these metrics primarily focus on static properties of saliency maps, such as overlap with annotations or explicit calculation of precision. They fail to assess the stability and consistency of explanations across varying thresholds for visualization. A more comprehensive approach that aggregates these metrics across all possible thresholds is needed.

In addressing these identified challenges, the proposed method effectively handles high-resolution images through multiscale adaptability, facilitating fine-grained localisation while incorporating robust qualitative evaluation metrics. This approach enhances interpretability and supports broader clinical applicability of XAI in medical imaging.

\section{Materials and Methods}

\subsection{Data Collection}
Our study used TMA samples from a randomised radiotherapy clinical trial at St George Hospital, Sydney, Australia (ClinicalTrials.gov NCT00138814) \cite{10.3390/cancers12123749}. Each patient's tumour was sampled with a $3\times1$ mm core using the Beecher Manual Arrayer MTA-1, as previously described \cite{10.3390/cancers12123749}. Ethics approval was provided by the South East Sydney Local Health District (SESLHD) Human Research Ethics Committee (HREC) at the Prince of Wales Hospital (PoWH), Sydney, Australia (HREC 96/16). From this TMA dataset, we used 140 tumour cores for training and 53 cores with pathologist annotations for testing. Additionally, we used four benign WSIs, from which we created 140 benign samples for classifier training. Ethics approval for benign breast slides was obtained from the HREC committee at the PoWH (HREC/17/POWH/389-17/176).

\begin{figure}[!t]
\centering
\includegraphics[width=\textwidth]{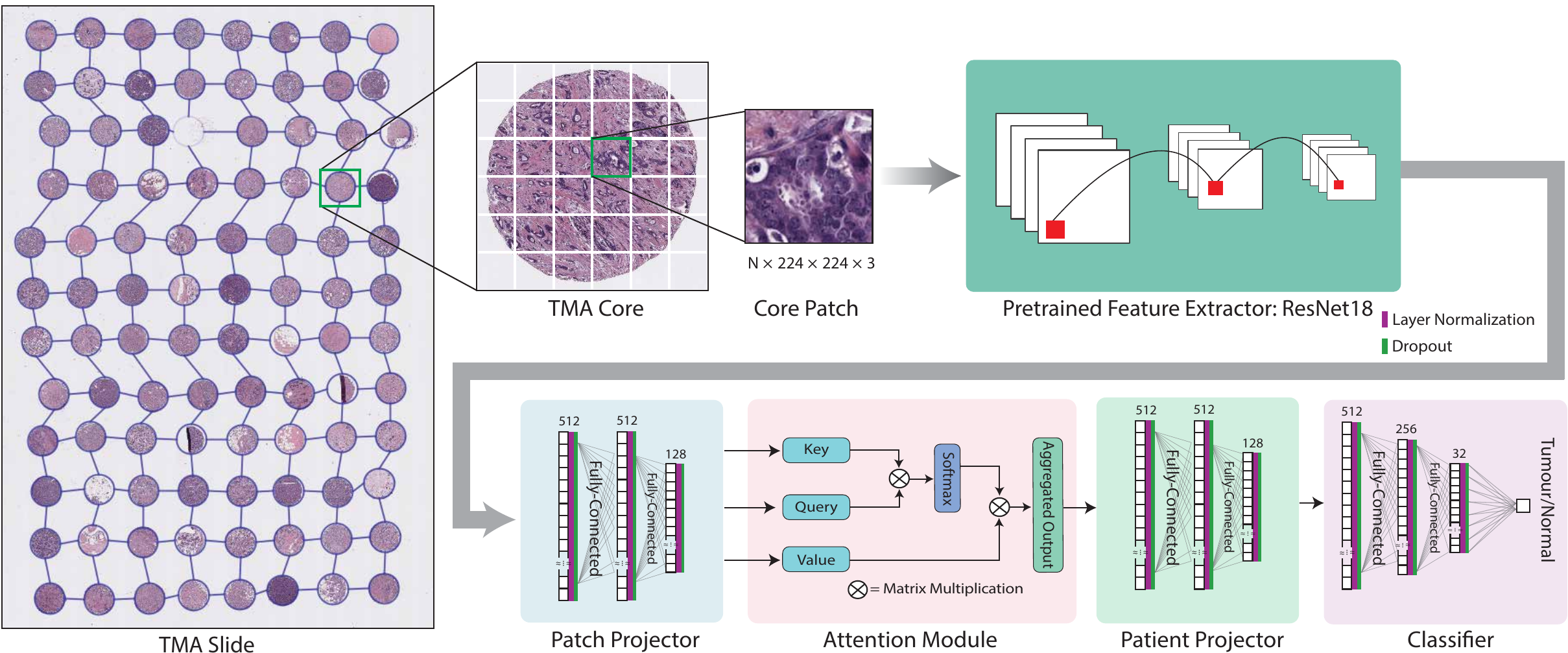}
\caption{Architecture of the TMA classification model (tumour versus normal). The pipeline begins with a TMA slide, which contains multiple circular TMA cores. Each core is divided into $N$ patches, each $224 \times 224 \times 3$ pixels. The patches from each core are processed using a ResNet18 model pretrained on the NCT-CRC-HE-100K dataset with 100,000 non-overlapping patches from H\&E-stained colorectal cancer and normal tissues spanning 9 classes. The attention module aggregates patch-level information at the core level, which is passed to the patient projector. The final classification is made through a dense layer classifier, distinguishing between tumour and normal tissue.}
\label{fig:trainingclassifier}
\end{figure}   

\subsection{Data Preparation}

\subsubsection{Slide Annotation}
An expert breast pathologist manually annotated the 53 selected TMA core slides using QuPath \cite{10.1038/s41598-017-17204-5}. The annotations outlined tumour regions, excluding necrotic areas while including stroma and tumour-infiltrating lymphocytes (TILs). The pathologist was blinded to molecular and clinical features during the annotation process, ensuring unbiased localisation of tumour regions for subsequent analysis.

\subsubsection{Image Preprocessing}
We obtained hematoxylin and eosin (H\&E)-stained tissue samples from both the TMA dataset and the benign WSIs. To prepare the images for analysis, we employed a semi-automated approach using QuPath software to create tissue masks that excluded artifacts and non-tissue regions. The annotated tumour areas were then segmented into non-overlapping patches of size $224\times224$ pixels, yielding about 500 patches per sample. To address variations in staining between samples, we applied vector-based colour normalisation following the method proposed by Macenko et al.~\cite{10.1109/ISBI.2009.5193250}. These preprocessing steps were crucial for generating high-quality image patches suitable for downstream analysis.

\subsection{Proposed Methods}
We propose a novel framework that generates interpretable saliency maps for digital pathology. The framework consists of two key components: (1) a MIL-based classifier designed for accurate tissue classification, and (2) GRAPHITE, a novel graph-based approach that integrates multiscale information to highlight diagnostically significant regions in TMAs. We describe these two components in detail.

\subsubsection{Stage 1: MIL-Based Classification}
In the first stage of the framework (Figure \ref{fig:trainingclassifier}), we focus on developing a classification model that distinguishes between tumour and normal tissue using high-resolution patches extracted at maximum magnification. In MIL for histopathology, each tissue microarray core is treated as a bag containing multiple patch instances, where a bag is labeled positive if it contains at least one cancerous patch and negative if all patches are benign. This approach significantly reduces the annotation burden since pathologists only need to provide core-level labels rather than detailed patch-level annotations, making it more practical for analysing large-scale histopathology datasets \cite{6853873}. Our methodology implements this framework through the following steps:

\paragraph{\bf Patch Extraction}
TMAs are typically composed of circular cores, each representing a distinct region of interest. At 40$\times$ magnification, a spatial resolution of 0.25 \textmu m per pixel is achieved, which provides sufficient detail for histological analysis of cell structures, nuclei and other tissue features. Each core is systematically divided into non-overlapping patches, with each patch being $224 \times 224$ pixels in size, denoted as $\mathbf{x}_i \in \mathbb{R}^{224 \times 224 \times 3}$, where the third dimension represents the RGB channels from the H\&E staining. For a given TMA core $k$, the set of patches extracted is represented as $\mathcal{P}_k = \{\mathbf{x}_i\}_{i=1}^{N_k}$, where $N_k$ is the total number of patches extracted from core $k$. Each patch is treated as an independent data point, processed through the feature extraction pipeline.

\begin{figure}[!t]
\includegraphics[width=\textwidth]{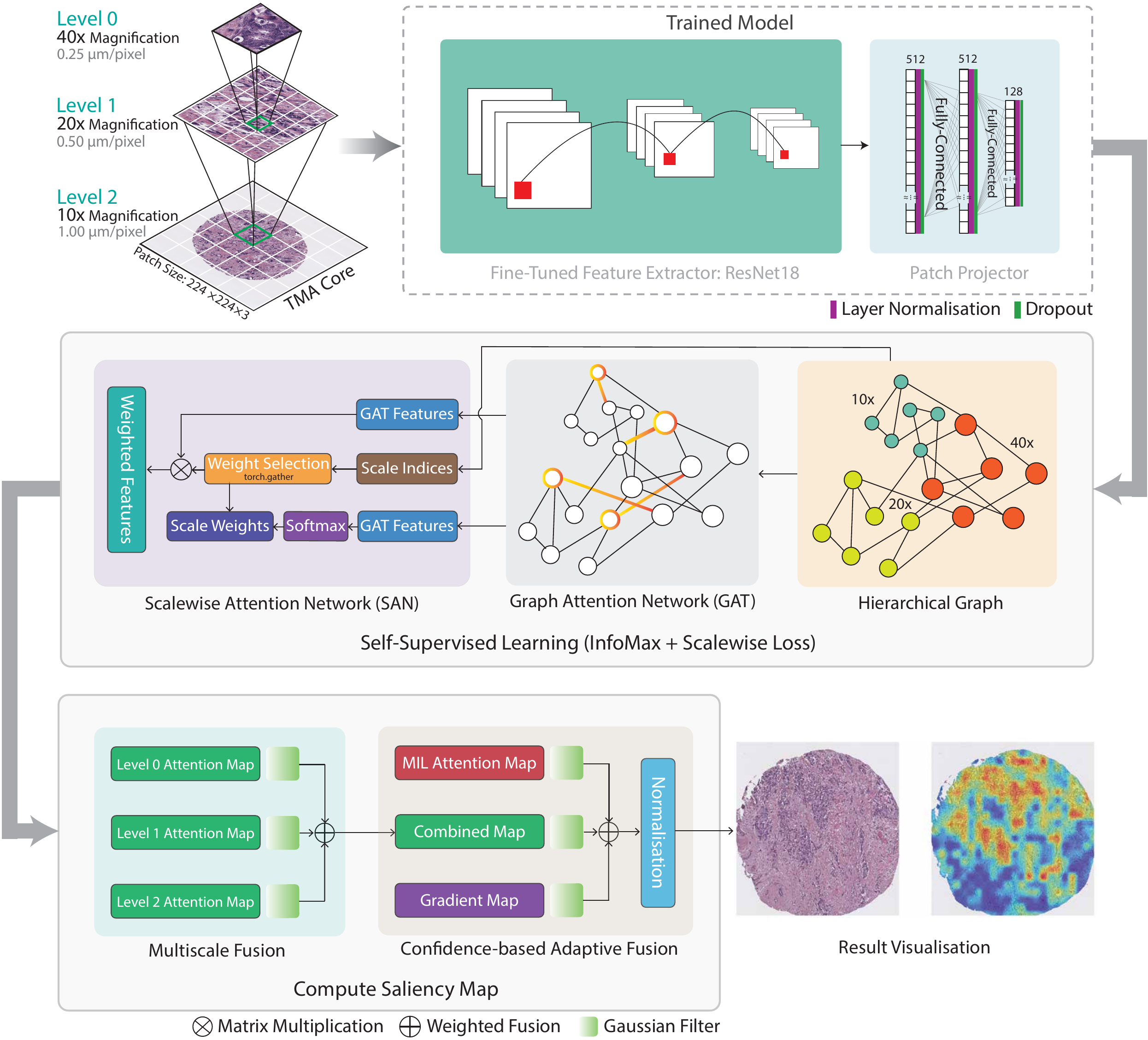}
\caption{Multiscale hierarchical model for saliency mapping in TMA analysis. Core patches are processed at three magnification levels: Level 0 (40$\times$, 0.25 \textmu m/pixel), Level 1 (20$\times$, 0.50 \textmu m/pixel) and Level 2 (10$\times$, 1 \textmu m/pixel) with 224$\times$224-pixel patches. These are passed through a fine-tuned ResNet18 for feature extraction and structured into an hierarchical graph, where edges represent spatial relationships. A graph attention network (GAT) applies attention weights, and a scalewise attention network (SAN) integrates multiscale information. Finally, a saliency map is computed through multilevel weighted and confidence-based fusion, highlighting key tumour regions for diagnostic interpretation.}
\label{fig:architecture}
\end{figure}  

\paragraph{\bf Feature Extraction With ResNet18}
ResNet18, a deep CNN, is used as the feature extractor in the first stage of the pipeline \cite{resnet, Pocock2022}. The model we use is pretrained on the \href{https://huggingface.co/datasets/1aurent/NCT-CRC-HE}{NCT-CRC-HE-100K} dataset, which includes 100,000 patches of colorectal cancer and normal tissue. The pretrained weights\footnote{https://huggingface.co/1aurent/resnet18.tiatoolbox-kather100k} provide a strong initialisation, allowing the model to capture general histological features.

For each patch $\mathbf{x}_i$, the ResNet18 model computes a feature representation $\mathbf{h}_i \in \mathbb{R}^D$, where $D$ denotes the dimension of the extracted feature vector. This feature vector encapsulates the hierarchical features learned by the CNN, from low-level textures and edges to high-level tissue structures:
\[
\mathbf{h}_i = f_{\text{ResNet18}}(\mathbf{x}_i)
\]
Here, $f_{\text{ResNet18}}: \mathbb{R}^{224 \times 224 \times 3} \rightarrow \mathbb{R}^D$ denotes the mapping from the raw image patch to the feature space. The dimension $D$ is typically 512 for ResNet18, capturing a rich set of features that summarise the histological information within each patch. The feature vectors are then used in downstream processes to determine the classification of each TMA core.

\paragraph{\bf Patch-Level Aggregation Using Attention Mechanism}
After features are extracted from each patch, they are first transformed into an embedding space using a patch projector. The patch projector consists of a series of fully connected layers (512 and 128 units) that project the high-dimensional patch features into a more compact and discriminative embedding space. This embedding serves as the foundation for further processing in both classification and subsequent visualisation stages.

Once the patch features are projected into the embedding space, they are aggregated into a core-level representation using a self-attention mechanism which allows the model to assign an importance weight $\alpha_i$ to each patch, reflecting its relevance to the overall classification task. The attention mechanism operates by computing query, key and value vectors for each patch \cite{10.5555/3295222.3295349}. These vectors are used to compute the attention weights as follows:
\[
\alpha_i = \frac{\exp\left( \mathbf{q}_i^\top \mathbf{k}_i / \sqrt{d_k} \right)}{\sum_{j=1}^{N_k} \exp\left( \mathbf{q}_j^\top \mathbf{k}_j / \sqrt{d_k} \right)}
\]
where $\mathbf{q}_i$, $\mathbf{k}_i$, and $\mathbf{v}_i$ represent the query, key and value vectors for patch $i$, and $d_k$ is the dimension of the key vectors. The self-attention module computes a weighted sum of the value vectors based on these attention weights. The core-level feature vector $\mathbf{z}_k$ is then obtained as:
\[
\mathbf{z}_k = \sum_{i=1}^{N_k} \alpha_i \mathbf{v}_i
\]
This process ensures that patches that contribute the most to the classification task are assigned higher attention, enabling the model to focus on the most relevant patches in differentiating between tumour and normal regions. The use of self-attention at this stage allows the model to capture complex relationships between patches, further improving its ability to identify critical histological features.

\paragraph{\bf Core-Level Classification and Patient Projector}
Following the attention-based aggregation of patch-level features,  the core-level feature vector $\mathbf{z}_k$ representing the TMA core is passed through the patient projector, which consists of two fully connected layers (respectively 512 and 128 units). These layers further transform the core-level feature into a more compact and discriminative representation. The transformed feature vector $\mathbf{p}_j = f_{\text{dense}}(\mathbf{z}_k)$ is then passed to the final classifier, which performs binary classification by computing the probability $\hat{y}_k \in [0, 1]$ that the core $k$ contains tumour tissue. This probability is obtained by applying a sigmoid function $\zeta(\mathbf{W}^\top \mathbf{z}_k + b)$, where $\mathbf{W}$ and $b$ are the learned weights and bias of the dense layer. The model is trained by minimising the binary cross-entropy loss:
\[
\mathcal{L} = -\frac{1}{K} \sum_{k=1}^{K} \left( y_k \log(\hat{y}_k) + (1 - y_k) \log(1 - \hat{y}_k) \right)
\]
where $y_k$ is the ground truth label for core $k$ and $K$ is the total number of training cores. This process ensures that the final classification decision reflects the pathological state (tumour versus normal) of each core.

\subsubsection{Stage 2: GRAPHITE-Based Visualisation}
The second stage of the framework uses a novel visualisation technique named GRAPHITE (Figure \ref{fig:architecture}), which is designed to generate interpretable saliency maps for TMA analysis. GRAPHITE operates on TMA core patches across multiple magnification levels, incorporating hierarchical graph structures and attention mechanisms to produce saliency maps that highlight diagnostically significant regions within a tissue core. Here we provide an in-depth technical explanation of each component of the GRAPHITE model.

\paragraph{\bf Multiscale Patch Extraction}
GRAPHITE begins with the extraction of patches from TMA cores at multiple magnification levels, capturing tissue features at varying spatial resolutions. Three magnification levels are employed: Level 0 (40$\times$) at a resolution of 0.25 \textmu m/pixel, Level 1 (20$\times$) at a resolution of 0.50 \textmu m/pixel, and Level 2 (10$\times$) at a resolution of 1.00 \textmu m/pixel. For each magnification level, patches of size $224 \times 224$ pixels are extracted. These patches represent specific regions of interest within the TMA core. Higher magnifications capture detailed cellular-level features, while lower magnifications provide a broader contextual view of the surrounding tissue architecture. Let $\mathcal{P}^m_k = \{\mathbf{x}^m_i\}_{i=1}^{N^m_k}$ denote the set of patches extracted from core $k$ at magnification level $m$, where $N^m_k$ is the number of patches at magnification $m$.

\paragraph{\bf Feature Extraction Using Trained Model}
Patch-level features are extracted using a combination of ResNet18 and a patch projector. Each patch is first processed by ResNet18, which outputs a feature vector representing the patch in a high-dimensional feature space. Both the ResNet18 model and the patch projector are fine-tuned during the classification stage (Stage 1) to extract domain-specific histopathology features effectively. The patch-level features are then further refined using the patch projector, which transforms the ResNet18 output into a compact, discriminative embedding (128 dimensions). This embedding facilitates better integration of patch-level features across multiple magnification levels, enhancing the model’s ability to perform multiscale aggregation and analysis. These combined features from ResNet18 and the patch projector serve as the foundation for the subsequent stages of the GRAPHITE framework for improved visualisation.

\paragraph{\bf Hierarchical Graph Construction}
The key innovation in GRAPHITE is its use of an hierarchical graph structure to model the spatial and scalewise relationships between patches of the same core at different magnification levels. Each node in the graph corresponds to a patch, and edges between nodes represent spatial relationships, both within and across magnification levels. Formally, let $\mathcal{G} = (\mathcal{V}, \mathcal{E})$ represent the hierarchical graph, where $\mathcal{V}$ is the set of nodes and $\mathcal{E}$ is the set of edges. $\mathcal{V}$ is composed of patches $\mathcal{P}_k^m$ extracted from the TMA core $k$ at magnification level $m$. Thus, the overall set of nodes is defined as:
\[
\mathcal{V} = \bigcup_{m=0}^{M-1} \mathcal{P}_k^m
\]
where $M$ is the number of magnification levels. The set of edges  $\mathcal{E}$ includes two types of edges: intrascale edges, which connect spatially adjacent patches at the same magnification level, and interscale edges, which link patches across different magnification levels based on their spatial alignment. The intrascale edges are constructed by connecting nodes at the same level if their spatial distance, calculated by a normalised Euclidean distance:
\[
d_{\text{spatial}} = \frac{\sqrt{(x_1 - x_2)^2 + (y_1 - y_2)^2}}{224}
\]
is less than or equal to a predefined \textit{spatial threshold}, where \((x_1, y_1)\) and \((x_2, y_2)\) are the coordinates of the centres of two patches at the same level. If $d_{\text{spatial}} \leq \text{spatial threshold}$, an edge is created between the two nodes. The interscale edges connect patches across different magnification levels by identifying hierarchically related patches based on a \textit{scale threshold}. For two patches at different levels, say $i$ and $j$, with a level difference $\Delta = |i - j|$, the coordinates of the patch at the lower magnification level are scaled by a factor of $2^\Delta$ to match the scale of the higher magnification level. The two patches are connected if their distance after scaling: \[
d_{\text{scale}} = \sqrt{(x_1 \cdot 2^{\Delta} - x_2)^2 + (y_1 \cdot 2^{\Delta} - y_2)^2}
\]
falls within the scale threshold, where $(x_1, y_1)$ and $(x_2, y_2)$ are the coordinates of patches at different levels. If $d_{\text{scale}} \leq \text{scale threshold} \times 224$, an edge is created between them.

This hierarchical graph structure enables the model to capture both fine-grained spatial relationships within each magnification level and hierarchical dependencies across different magnifications. By aggregating patch-level information across scales and incorporating spatial context, the model can produce a more comprehensive and interpretable saliency map, effectively highlighting regions critical to its predictions. This multiscale approach is crucial for generating meaningful visual explanations that align with the structural and spatial intricacies of tissue morphology.

\paragraph{\bf Graph Attention Network (GAT)}
The GAT in GRAPHITE plays a crucial role in refining patch-level features by leveraging spatial and cross-scale relationships captured within the hierarchical graph. After constructing the hierarchical graph from patches extracted across multiple magnification levels, GAT aggregates information from neighbouring patches, enabling the model to capture both fine-grained spatial detail and broader contextual dependencies across scales. GAT is composed of two parallel attention mechanisms that process different types of edges: spatial edges that connect patches within the same magnification level, and cross-scale edges that link patches across different magnification levels (Figure \ref{fig:architecture}). Each mechanism employs multihead attention, allowing the model to assign distinct weights to neighbouring nodes and focus on the most relevant spatial and hierarchical relationships.

For each node \( v_i \in \mathcal{V} \), GAT computes an updated feature vector \( \mathbf{h}'_i \) by aggregating features from its spatial and cross-scale neighbours separately. The updated feature vector is computed as:
\begin{equation}
\mathbf{h}'_i = 
\sum_{j \in \mathcal{N}_{\text{spatial}}(i)} \psi_{ij}^{\text{spatial}} \mathbf{W}_{\text{spatial}} \mathbf{h}_j 
\ + \!\!\!
\sum_{j \in \mathcal{N}_{\text{cross}}(i)} \psi_{ij}^{\text{cross}} \mathbf{W}_{\text{cross}} \mathbf{h}_j
\end{equation}
where \( \mathcal{N}_{\text{spatial}}(i) \) and \( \mathcal{N}_{\text{cross}}(i) \) denote the spatial and cross-scale neighbours of node \( i \) respectively. In this hierarchical graph structure, spatial edges (\( t_{ij} = \text{spatial} \)) connect nodes at the same hierarchical level and represent local neighbourhood relationships, while cross-scale edges (\( t_{ij} = \text{cross} \)) connect nodes between different hierarchical levels, capturing multiscale interactions in the graph. \( \mathbf{W}_{\text{spatial}} \) and \( \mathbf{W}_{\text{cross}} \) are learnable weight matrices corresponding to spatial and cross-scale edges, while \( \psi_{ij}^{\text{spatial}} \) and \( \psi_{ij}^{\text{cross}} \) are the attention weights assigned to the spatial and cross-scale edges respectively. These attention weights are computed using the following equation:
\begin{equation}
\psi_{ij}^{t_{ij}} = 
\frac{\exp \left( \text{LeakyReLU}\left( \mathbf{a}_{t_{ij}}^\top \left[ \mathbf{W}_{t_{ij}} \mathbf{h}_i \, \| \, \mathbf{W}_{t_{ij}} \mathbf{h}_j \right] \right) \right)}{\displaystyle \sum_{n \in \mathcal{N}_{t_{ij}}(i)} \exp \left( \text{LeakyReLU}\left( \mathbf{a}_{t_{ij}}^\top \left[ \mathbf{W}_{t_{ij}} \mathbf{h}_i \, \| \, \mathbf{W}_{t_{ij}} \mathbf{h}_n \right] \right) \right)}
\end{equation}
where \( t_{ij} \in \{\text{spatial}, \text{cross}\} \) indicates the type of edge (spatial or cross-scale). Specifically, for spatial edges, \( \psi_{ij}^{\text{spatial}} \) is computed by substituting the corresponding parameters \( \mathbf{a}_{\text{spatial}} \) and \( \mathbf{W}_{\text{spatial}} \). Similarly, for cross-scale edges, \( \psi_{ij}^{\text{cross}} \) is computed using \( \mathbf{a}_{\text{cross}} \) and \( \mathbf{W}_{\text{cross}} \). Symbol \( \| \) denotes concatenation of the transformed feature vectors of nodes \( i \) and \( j \), and \(\text{LeakyReLU}\) is the activation function applied within the attention mechanism. The normalisation in the denominator ensures that the attention weights sum to 1 over the respective neighbourhood \( \mathcal{N}_{t_{ij}}(i) \). 

The multihead attention mechanism in GAT allows the model to attend to different aspects of neighbouring patches, assigning higher weights to those connections that are more informative for feature enhancement. By combining the features obtained from both spatial and cross-scale edges, the GAT module in GRAPHITE provides an enriched feature representation, which is crucial for accurate and interpretable visualisation of important regions within tissue samples.

\begin{figure}[!t]
\centering
\includegraphics[width=\textwidth]{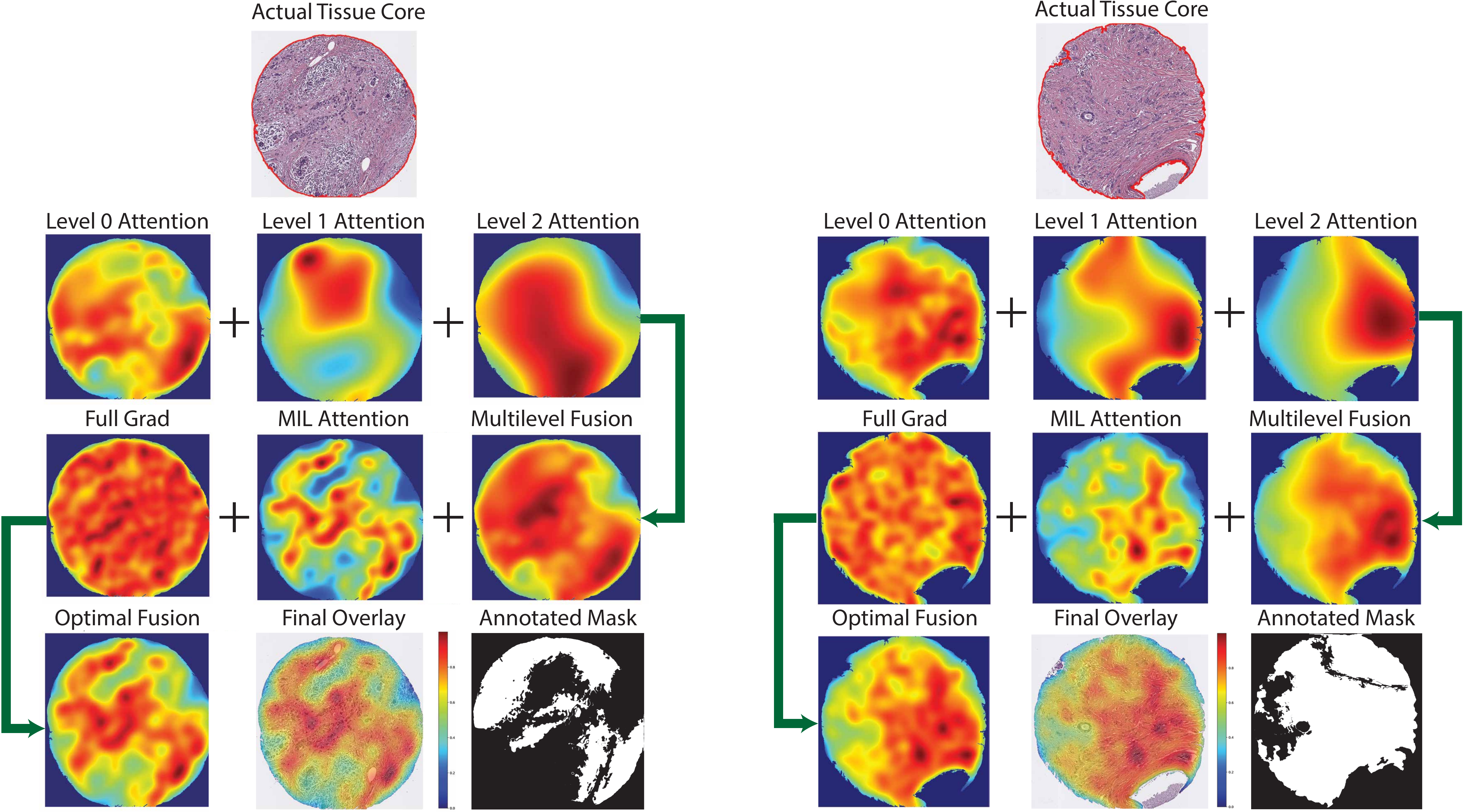}
\caption{Visualisation pipeline of GRAPHITE for breast cancer TMA analysis of two sample cores. Row 1: The actual tissue cores. Row 2: The attention maps across three levels (Level 0, Level 1, and Level 2), representing multiscale analysis with different magnification levels. Row 3: The attention maps are fused through a multilevel fusion process, after which attention maps are generated by FullGrad and MIL methods, which are integrated to enhance feature interpretability. Row 4: Confidence-based optimal fusion combines the multilevel attention maps, creating a refined representation of salient regions. The Final Overlay visually aligns this fused attention map with the tissue core, and the Annotated Mask indicates the pathologist-verified cancerous regions.}
\label{heatmap_fusion}
\end{figure}

\paragraph{\bf Scalewise Attention Network (SAN)}
Following the GAT-based feature enhancement, SAN is employed to learn importance weights specific to each magnification level, enabling the model to integrate multiscale information effectively. For each magnification level \( m \) in the hierarchy, SAN computes a level-specific attention score for patches at that level. The network consists of a set of attention modules, each dedicated to a particular magnification level, which outputs attention scores based on the node features.

For each patch \( i \) at magnification level \( m \), SAN computes an attention weight \( s^m_i \) as follows:
\[
s^m_i = \frac{\exp(a^m(\mathbf{h}^m_i))}{\sum_{m'=0}^{M-1} \exp(a^{m'}(\mathbf{h}^{m'}_i))}
\]
where \( a^m(\cdot) \) is a learnable function that assigns an importance score to the feature vector \( \mathbf{h}^m_i \). The softmax operation ensures that the attention weights across scales sum to 1, allowing the model to focus on the magnification levels that provide the most relevant information for each patch.

Once the level-specific attention scores are computed, SAN aggregates the features at each level to form a unified multiscale representation. The weighted feature vector for each level is given by:
\[
\mathbf{h}^m_{\text{weighted}} = s^m_i \cdot \mathbf{h}^m_i
\]
After obtaining the weighted features for all levels, SAN computes a cross-scale attention score to integrate information across scales. This cross-scale attention mechanism assigns an additional weight to each level output, enabling further refinement of the multiscale representation.

The final integrated feature vector for patch \( i \), denoted as \( \mathbf{h}^{\text{multi}}_i \), is computed as:
\[
\mathbf{h}^{\text{multi}}_i = \sum_{m=0}^{M-1} c^m \mathbf{h}^m_{\text{weighted}}
\]
where \( c^m \) represents the cross-scale attention weight for each level \( m \), which is learned through a separate cross-scale attention network. By dynamically assigning weights at both the individual and cross-scale levels, SAN ensures that the final patch representation \( \mathbf{h}^{\text{multi}}_i \) emphasises the most relevant scales, providing a comprehensive multiscale feature representation for the visualisation task.

\paragraph{\bf Saliency Map Generation Using Multiscale and Confidence-Based Fusion}
In GRAPHITE, the saliency map generation employs the fusion of attention maps using multiple strategies, including multilevel fusion and confidence-based adaptive fusion. This results in a final saliency map that integrates relevant information from different attention mechanisms (Figure~\ref{heatmap_fusion}).

The first step involves multilevel fusion, where attention maps from different levels (Level 0, Level 1, and Level 2) are weighted and combined to capture features at various scales. The attention map at each level provides unique insights, with higher-resolution levels capturing finer details, while lower-resolution levels capture broader contextual information. To emphasize the importance of fine details, each level is assigned a weight determined empirically: Level 0 is given the highest weight of \( \rho_0 = 0.5 \), Level 1 a moderate weight of \( \rho_1 = 0.3 \), and Level 2 a lower weight of \( \rho_2 = 0.2 \). Each map is smoothed using a Gaussian filter with different values of \(\sigma\) (1, 2, and 4, respectively) to enhance spatial coherence at each scale. The multilevel fusion is expressed as:
\[
M_{\text{combined}} = \rho_0 \cdot \text{Gaussian}(A_0, \sigma=1) + \rho_1 \cdot \text{Gaussian}(A_1, \sigma=2) + \rho_2 \cdot \text{Gaussian}(A_2, \sigma=4)
\]
where \(A_0\), \(A_1\), and \(A_2\) are the attention maps at the respective level. This weighted fusion creates a combined map \(M_{\text{combined}}\) that prioritises fine-grained information from higher-resolution levels while integrating contextual information from lower resolutions, resulting in a balanced representation of important features across scales.

Next, method-specific enhancements are applied to the MIL attention map and the FullGrad gradient map. A Gaussian filter is applied to each map, with \(\sigma=2\) for the MIL map to highlight tissue patterns and \(\sigma=1\) for the FullGrad map to capture local features. This produces enhanced versions of each map, emphasising their unique strengths.

To adaptively weight these maps, confidence scores are calculated for each component based on the prominence of high-activation regions. The confidence score for each map is defined as the mean of values in the top tenth percentile of activations. This score calculation is represented as:
\[
C = \text{mean}(F[F > \text{percentile}(F, 90)])
\]
where \(F\) refers to each of the combined or enhanced maps (\(F_{\text{combined}}\), \(F_{\text{MIL}}\) and \(F_{\text{FullGrad}}\)). These confidence scores—\( C_{\text{combined}} \), \( C_{\text{MIL}} \) and \( C_{\text{FullGrad}} \)—are then normalised to determine the adaptive weights (\( v \)) for each map, ensuring that more confident maps contribute more to the final result. The weights are computed as:
\[
C_{\text{total}} = C_{\text{combined}} + C_{\text{MIL}} + C_{\text{FullGrad}}
\]
\[
v_{\text{combined}} = 0.6 \cdot \frac{C_{\text{combined}}}{C_{\text{total}}}, \quad v_{\text{MIL}} = 0.3 \cdot \frac{C_{\text{MIL}}}{C_{\text{total}}}, \quad v_{\text{FullGrad}} = 0.1 \cdot \frac{C_{\text{FullGrad}}}{C_{\text{total}}}
\]

Finally, the confidence-based fusion combines the multiscale, MIL, and FullGrad maps as follows:
\[
S_{\text{fused}} = v_{\text{combined}} \cdot M_{\text{combined}} + v_{\text{MIL}} \cdot M_{\text{MIL}} + v_{\text{FullGrad}} \cdot M_{\text{FullGrad}}
\]

The resulting fused map is then normalised to ensure the values are within a suitable range for visualization:
\[
S_{\text{fused}} = \frac{S_{\text{fused}} - \min(S_{\text{fused}})}{\max(S_{\text{fused}}) - \min(S_{\text{fused}}) + \text{1e-8}}
\]
where 1e-8 is added to prevent division by zero during the normalisation process. This adaptive fusion process ensures that the final saliency map captures multiscale attention information, method-specific strengths and confidence-based sensitivity, providing a detailed and interpretable visualisation of regions critical to the model predictions.

\subsection{Model Training}
The training of the proposed GRAPHITE model consists of two stages. The first stage involves training a classifier in a MIL framework, distinguishing between positive bags (containing both tumour and normal regions) and negative bags (containing only normal regions). The second stage leverages self-supervised learning with a combination of InfoMax and scalewise loss to further refine the feature representations for multiscale aggregation.

\subsubsection{Stage 1: Classifier Training With MIL}
In the training at Stage 1, we utilise a pretrained ResNet18 model as a feature extractor to process patches from the TMA cores (Figure~\ref{fig:trainingclassifier}). Each core is divided into patches, which are passed through the ResNet18 backbone to obtain feature vectors for each patch. These patch-level features are subsequently projected through a series of fully connected layers (the Patch Projector) to generate a lower-dimensional representation.

In a MIL setting, patches are grouped into bags, where a positive bag contains both tumour and normal patches, and a negative bag consists solely of normal patches. The model is trained to classify bags rather than individual patches, focusing on the detection of tumour regions within positive bags. During training, the patch-level features are aggregated through attention mechanisms to produce a bag-level representation, which is then fed into a binary classifier that learns to differentiate between positive and negative bags, identifying regions that are indicative of tumour presence.

The classifier is optimised using a binary cross-entropy loss with the Adam optimiser \cite{mao2023crossentropylossfunctionstheoretical, kingma2017adammethodstochasticoptimization}. We used a learning rate of 0.001 and early stopping with a patience of 4 epochs to prevent overfitting. Each training run was conducted with a minibatch of 12 bags per step, continuing until the early stopping criterion was met or the maximum of 150 epochs was reached. The best model weights, based on validation performance, were stored for use in subsequent stages.

\subsubsection{Stage 2: Self-Supervised Training With InfoMax and Scalewise Loss}
In the second stage, the model undergoes self-supervised training to further enhance feature representations through multiscale learning (Figure~\ref{fig:architecture}). This stage combines two objectives: InfoMax loss and scalewise loss, which together help the model capture both global and scale-specific information effectively.

\paragraph{\bf InfoMax Loss} This loss is designed to maximise the mutual information between patch-level features and the core-level (or graph-level) representations. For each magnification level, this loss computes the similarity between the normalised embeddings of patches and the aggregated core-level embedding, promoting feature consistency across different scales. The InfoMax loss is calculated as:
\[
\mathcal{L}_{\text{InfoMax}} = -\frac{1}{| \mathcal{V} |} \sum_{i \in \mathcal{V}} \log \left( \frac{\exp(\text{sim}(\mathbf{h}_i, \mathbf{g}) / \tau)}{\exp(\text{sim}(\mathbf{h}_i, \mathbf{g}) / \tau) + \sum_{j \neq i} \exp(\text{sim}(\mathbf{h}_i, \mathbf{h}_j) / \tau)} \right)
\]
where \( \mathbf{h}_i \) represents the feature vector of node \( i \) and \( \mathbf{g} \) is the core-level (graph) representation. Here, \( \text{sim}(\cdot) \) denotes similarity, \( \tau \) is a temperature parameter, and the loss encourages the features of each node to be close to the core representation while also distinct from other node features.

\paragraph{\bf Scalewise Loss} This loss focuses on maintaining consistency and relevance across different magnification levels. It consists of two main components: interscale consistency and intrascale discrimination. Interscale consistency enforces similarity between embeddings from adjacent scales, calculated using the dot product between normalised embeddings of nodes from different scales. Intrascale discrimination aims to differentiate embeddings within the same scale by discouraging similarity among different nodes at the same level. For node embeddings \( \mathbf{h}_i \) and \( \mathbf{h}_j \) at levels \( m \) and \( l \), the scalewise loss is defined as:
\[
\mathcal{L}_{\text{scale}} = -\sum_{m=0}^{M-1} \sum_{l=m+1}^{M-1} w_m w_l \cdot \log \left( \frac{\exp(\text{sim}(\mathbf{h}_i^m, \mathbf{h}_j^l) / \tau)}{\exp(\text{sim}(\mathbf{h}_i^m, \mathbf{h}_j^l) / \tau) + \text{1e-8}} \right)
\]
where \( w_m \) and \( w_l \) are attention weights assigned to magnification levels \( m \) and \( l \), \( \tau \) is a temperature parameter, and a small constant to prevent numerical instability. This loss encourages the model to maintain consistency across scales while distinguishing nodes within the same scale.

\paragraph{\bf Total Loss} The total loss for self-supervised training in Stage 2 is defined as the sum of two losses:
\[
\mathcal{L}_{\text{total}} = \mathcal{L}_{\text{InfoMax}} + \mathcal{L}_{\text{scale}}
\]
This formulation assigns equal importance to both components, allowing GRAPHITE to effectively learn robust multiscale representations. These representations enhance the interpretability of the model and improve the accuracy of the generated saliency maps.

In this stage, the model is optimised using the Adam optimiser with a learning rate of 0.0005, early stopping with a patience of 4 epochs is applied to avoid overfitting, and training continues until the stopping criterion is met or for a maximum of 100 epochs. The final model weights from this stage are stored for visualisation.

\subsection{Evaluation Metrics}
To assess the effectiveness and interpretability of GRAPHITE and other XAI methods, we employed several evaluation metrics that capture different aspects of model performance and explainability. These metrics include the area under the receiver operating characteristic curve (AUROC), area under the precision-recall curve (AUPRC) \cite{10.1186/s13637-014-0012-3,10.1093/neuonc/noaa177,10.1016/j.jclinepi.2015.02.010,10.1111/2041-210x.13140}, mean average precision (mAP) \cite{10.1016/j.asoc.2021.108261,10.1109/TPAMI.2015.2437384}, threshold stability (ThS), threshold robustness (ThR), mean intersection over union (mIoU) \cite{10.1007/s11263-021-01533-0}, balanced accuracy (BA) \cite{10.1016/j.cmpb.2012.11.003}, comprehensive XAI performance score (CXPS) and the net benefit curve (NBC) \cite{10.3390/healthcare11162244}. Each metric is discussed in detail below, along with accompanying mathematical formulations.

\subsubsection{Area Under the Receiver Operating Characteristic Curve (AUROC)}
AUROC summarises the ability of the classifier to distinguish between positive and negative classes at  different cut-off thresholds. The receiver operating characteristic (ROC) curve plots the true-positive rate (TPR) against the false-positive rate (FPR):
\begin{equation}
    \text{TPR} = \frac{\text{TP}}{\text{TP} + \text{FN}}
\end{equation}
\begin{equation}
    \text{FPR} = \frac{\text{FP}}{\text{FP} + \text{TN}}
\end{equation}
where TP, FN, FP and TN represent the numbers of true positives, false negatives, false positives and true negatives respectively. The AUROC score ranging from 0 to 1 represents the area under the ROC curve, with higher AUROC values indicating better discrimination.

\subsubsection{Area Under the Precision-Recall Curve (AUPRC)}
AUPRC emphasises classifier performance when handling imbalanced data, as it focusses on the precision and recall of positive instances. The precision-recall curve plots precision as a function of recall:
\begin{equation}
    \text{Precision} = \frac{\text{TP}}{\text{TP} + \text{FP}}
\end{equation}
\begin{equation}
    \text{Recall} = \frac{\text{TP}}{\text{TP} + \text{FN}}
\end{equation}
The area under this curve is a robust indicator of performance, particularly in scenarios with skewed class distributions.

\subsubsection{Mean Average Precision (mAP)}
The mAP is a commonly used metric in information retrieval and object detection, reflecting both precision and recall across multiple thresholds. It is calculated by taking the average of the precision values at each relevant recall threshold, providing a single score to evaluate model accuracy and confidence. Denoting the precision and recall at the $n$th threshold by $P_n$ and $R_n$, we define average precision (AP) as:
\begin{equation}
    \text{AP} = \sum_{n} (R_n - R_{n-1}) P_n
\end{equation}
and the mAP is then computed as:
\begin{equation}
    \text{mAP} = \frac{1}{N} \sum_{i=1}^{N} \text{AP}_i
\end{equation}
where $N$ is the number of classes or models being evaluated. A higher mAP indicates better model performance in distinguishing positive from negative cases.

\begin{table}[!t]
\centering
\caption{Performance comparison of different XAI methods.}
\resizebox{0.8\textwidth}{!}{%
\begin{tabular}{l*{8}{c}}
\toprule
\textbf{Method} & \textbf{mAP} & \textbf{AUROC} & \textbf{AUPRC} & \textbf{mIoU} & \textbf{ThS} & \textbf{ThR} & \textbf{BA} & \textbf{CXPS} \\ 
\midrule
GradCAM       & 0.44 & 0.86 & 0.55 & 0.24 & 0.17 & 0.20 & 0.60 & 0.48 \\
GradCAM++     & 0.50 & 0.89 & 0.62 & 0.31 & 0.31 & 0.40 & 0.63 & 0.55 \\
AblationCAM   & 0.44 & 0.77 & 0.46 & 0.25 & 0.20 & 0.10 & 0.59 & 0.45 \\
EigenCAM      & 0.45 & 0.80 & 0.54 & 0.20 & 0.11 & 0.10 & 0.58 & 0.44 \\
FullGrad      & 0.52 & 0.91 & 0.65 & 0.34 & 0.36 & 0.50 & 0.64 & 0.58 \\
Attention     & 0.55 & 0.92 & 0.77 & 0.38 & 0.43 & 0.60 & 0.67 & 0.62 \\
SHAP         & 0.36 & 0.61 & 0.31 & 0.15 & 0.07 & 0.10 & 0.52 & 0.35 \\
LIME        & 0.45 & 0.83 & 0.52 & 0.25 & 0.15 & 0.30 & 0.60 &0.48 \\
GRAPHITE-Base & 0.55 & 0.93 & 0.77 & 0.37 & 0.40 & 0.60 & 0.67 & 0.62 \\
GRAPHITE-V1  & 0.55 & 0.93 & 0.76 & 0.39 & 0.47 & 0.60 & 0.68 & 0.63 \\
GRAPHITE-V2  & \textbf{0.56} & \textbf{0.94} & \textbf{0.78} & \textbf{0.41} & \textbf{0.50} & \textbf{0.70} & \textbf{0.68} & \textbf{0.65} \\
\bottomrule
\end{tabular}%
}
\label{tab:performance-comparison}
\end{table}

\subsubsection{Threshold Stability (ThS)}
ThS measures the consistency of  model performance across different thresholds, providing insight into its robustness in real-world scenarios. A high ThS score indicates that the model’s precision-recall balance is stable across a range of thresholds, which is essential for applications where decision thresholds might vary.

To calculate ThS, we use the following formula based on the F1 score (the harmonic mean of precision and recall):
\begin{equation}
    \text{F1} = 2 \cdot \frac{\text{Precision} \cdot \text{Recall}}{\text{Precision} + \text{Recall}}
\end{equation}
The ThS score is then calculated as:
\begin{equation}
    \text{ThS} = 1 - \frac{\sigma(\text{F1})}{\mu(\text{F1})}
\end{equation}
where $\sigma(\text{F1})$ and $\mu(\text{F1})$ denote the standard deviation and mean of F1 scores across different thresholds respectively. This formulation allows us to quantify model stability, with higher ThS values indicating more consistent performance across thresholds.

\subsubsection{Threshold Robustness (ThR)}
ThR evaluates the range of thresholds over which the model's F1 score remains within 95\% of its peak performance. This metric provides an estimate of how sensitive the model is to threshold changes, which is crucial for maintaining reliable predictions under varying conditions.

First, we calculate the F1 score as above. We then identify the peak F1 score ($\text{F1}_{\text{peak}}$) and determine the range of thresholds for which the F1 score is at least 95\% of this peak value:
\begin{equation}
    \text{ThR} = \text{max}(\text{threshold}) - \text{min}(\text{threshold}), \quad \text{where } \text{F1} \geq 0.95 \cdot \text{F1}_{\text{peak}}.
\end{equation}
If no thresholds meet this condition, ThR is set to zero. A larger ThR value indicates that the model maintains high performance across a wider range of thresholds, reflecting its robustness to threshold variation.

\subsubsection{Mean Intersection over Union (mIoU)}
The mIoU measures the overlap between the predicted and actual regions of interest in an image, commonly used in segmentation tasks. It is defined as:
\begin{equation}
    \text{IoU} = \frac{|A \cap B|}{|A \cup B|}
\end{equation}
where $A$ is the predicted region and $B$ is the ground truth region. The mean IoU across all samples is:
\begin{equation}
    \text{mIoU} = \frac{1}{N} \sum_{i=1}^{N} \text{IoU}_i
\end{equation}
Higher mIoU values indicate better alignment of the model's prediction with ground truth regions.

\subsubsection{Balanced Accuracy (BA)}
The BA metric accounts for both sensitivity and specificity in binary classification, offering a balanced metric that is not biased by class imbalance. It is defined as the mean of the true-positive rate (TPR, equal to sensitivity or recall) and true-negative rate (TNR, also called specificity):
\begin{equation}
    \text{BA} = \frac{\text{TPR} + \text{TNR}}{2}
\end{equation}

\subsubsection{Comprehensive XAI Performance Score (CXPS)}
The CXPS is a weighted composite metric developed to assess both the predictive accuracy and interpretability of XAI methods across several critical metrics. CXPS combines mAP, AUROC, mIoU, ThS, ThR and BA to provide a holistic measure of model performance and explainability. It is calculated as:
\begin{equation}
    \text{CXPS} = q_{\text{mAP}} \cdot \text{mAP} + q_{\text{AUROC}} \cdot \text{AUROC} + q_{\text{mIoU}} \cdot \text{mIoU} + q_{\text{ThS}} \cdot \text{ThS} + q_{\text{ThR}} \cdot \text{ThR} + q_{\text{BA}} \cdot \text{BA},
\end{equation}
where the weights \( q_{\text{mAP}} = 0.20 \), \( q_{\text{AUROC}} = 0.25 \), \( q_{\text{mIoU}} = 0.20 \), \( q_{\text{ThS}} = 0.10 \), \( q_{\text{ThR}} = 0.10 \) and \( q_{\text{BA}} = 0.15 \) reflect the relative importance of each metric. These weights are chosen to prioritise \( q_{\text{AUROC}} \) as the primary metric for evaluating the model’s discriminative ability, while \( q_{\text{mAP}} \) and \( q_{\text{mIoU}} \) focus on precision-recall balance and interpretability, respectively, \( q_{\text{ThS}} \) and \( q_{\text{ThR}} \) capture the model’s stability and robustness across thresholds which are essential for consistent performance, and \( q_{\text{BA}} \) ensures balanced accuracy, which is particularly valuable in imbalanced datasets. This weighted metric provides a balanced assessment of XAI methods, considering both performance and interpretability, making it suitable for selecting models in clinical applications where reliability and explainability are paramount.

\begin{figure}[!t]
\centering
\includegraphics[width=0.7\textwidth]{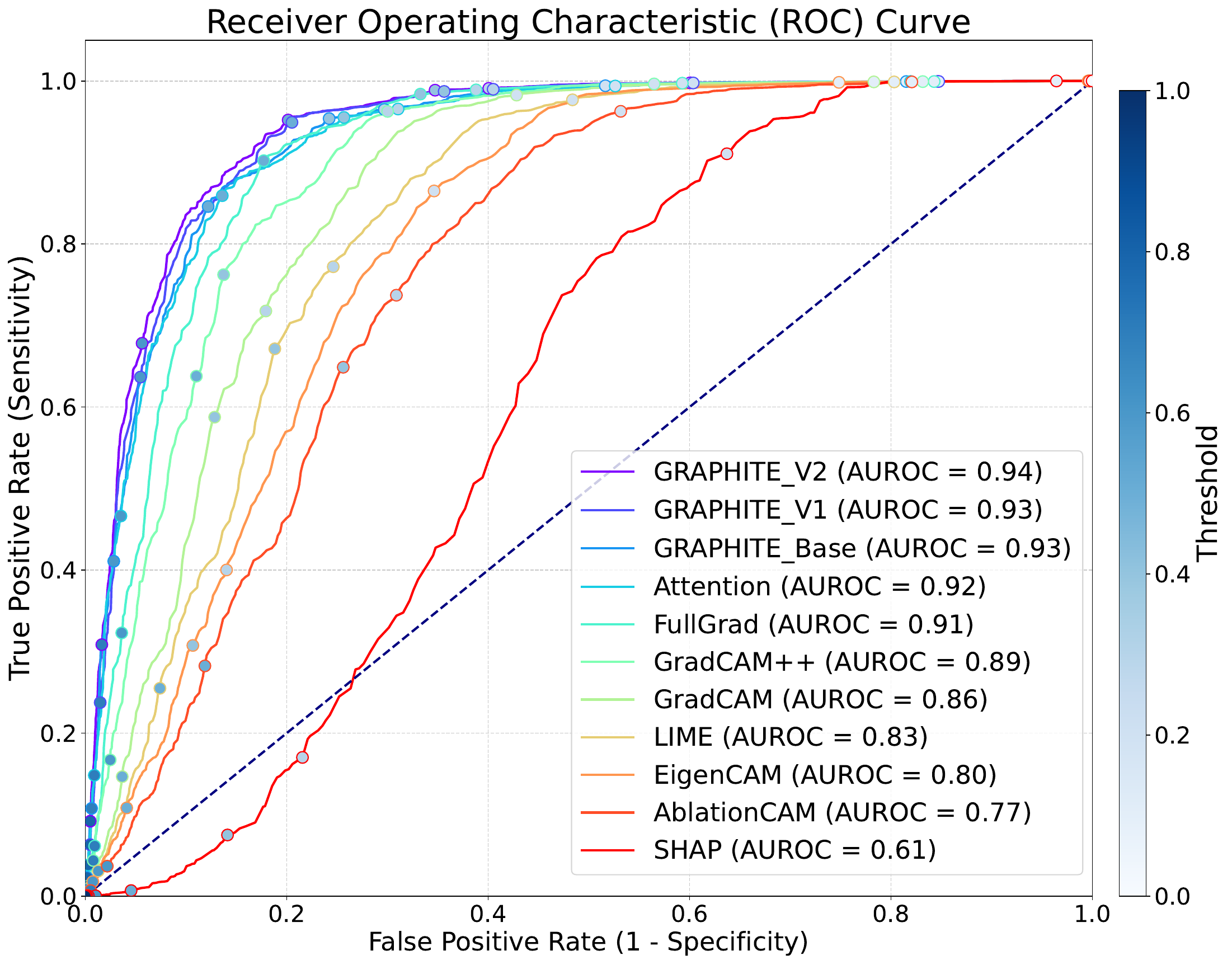}
\caption{Receiver operating characteristic (ROC) curves for various XAI methods on breast cancer TMA analysis. AUROC for each method is displayed in the legend, with GRAPHITE-V2 achieving the highest AUROC of 0.94, followed closely by GRAPHITE-V1 and GRAPHITE-Base, both at 0.93. This performance demonstrates the superior discriminative ability of GRAPHITE variants compared to other XAI methods, such as Attention (AUROC = 0.92) and FullGrad (AUROC = 0.91). Model-agnostic methods LIME (AUROC = 0.83) and SHAP (AUROC = 0.61) show lower performance in this context. The colour gradient along the curves represents different threshold values.}
\label{auroc}
\end{figure}

\begin{figure}[!t]
\centering
\includegraphics[width=0.7\textwidth]{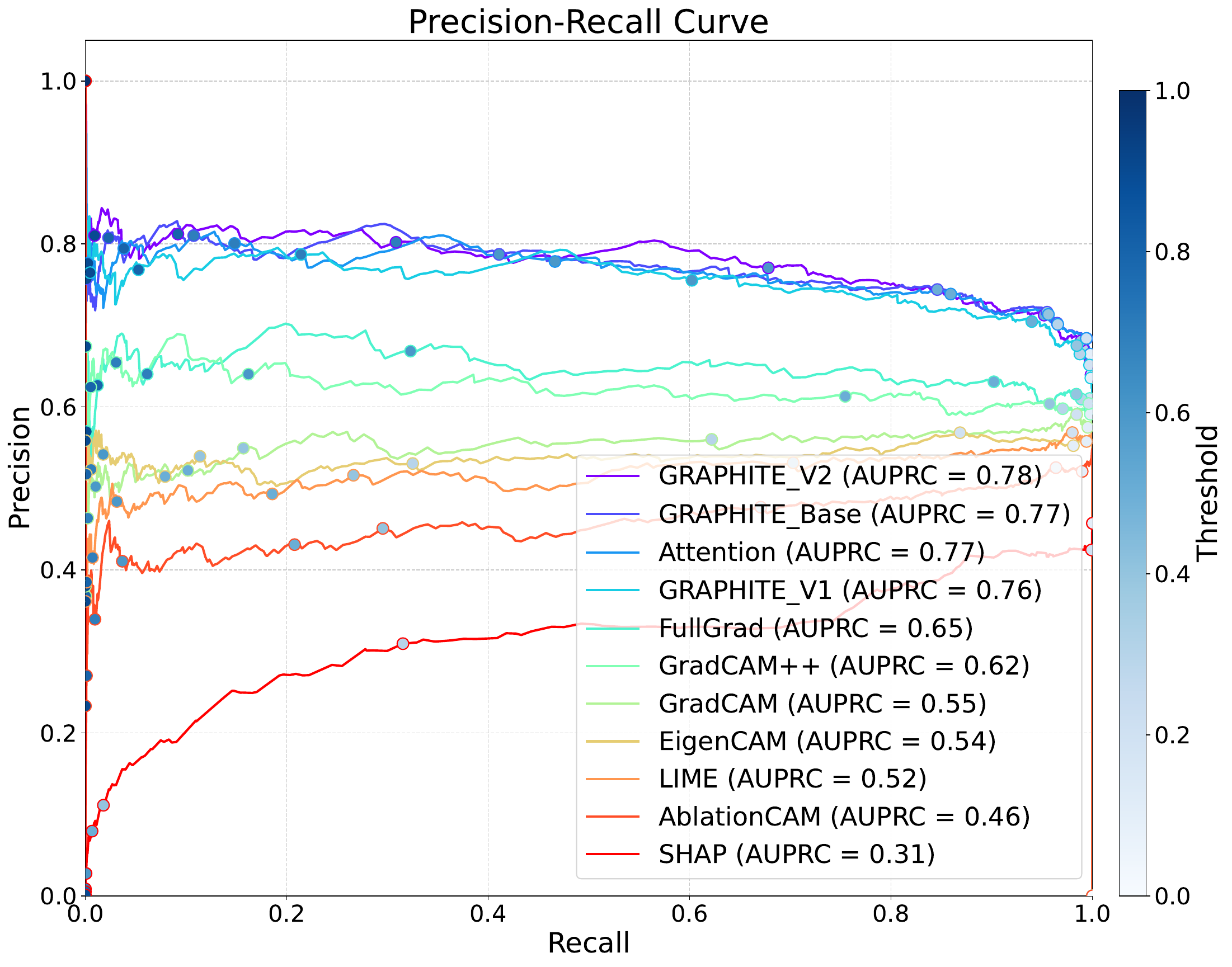}
\caption{Precision-recall (PR) curves for various XAI methods on breast cancer TMA analysis. AUPRC for each method is displayed in the legend, with GRAPHITE-V2 achieving the highest AUPRC of 0.78, followed by GRAPHITE-Base and Attention, both with an AUPRC of 0.77. This performance highlights the superior PR balance of GRAPHITE variants compared to other XAI methods, such as FullGrad (AUPRC = 0.65) and Grad-CAM++ (AUPRC = 0.62). LIME (AUPRC = 0.52) and SHAP (AUPRC = 0.31) exhibit significantly lower precision-recall performance. The colour gradient along the curves represents different threshold values.}
\label{auprc}
\end{figure}  

\begin{figure}[!t]
\centering
\includegraphics[width=0.8\textwidth]{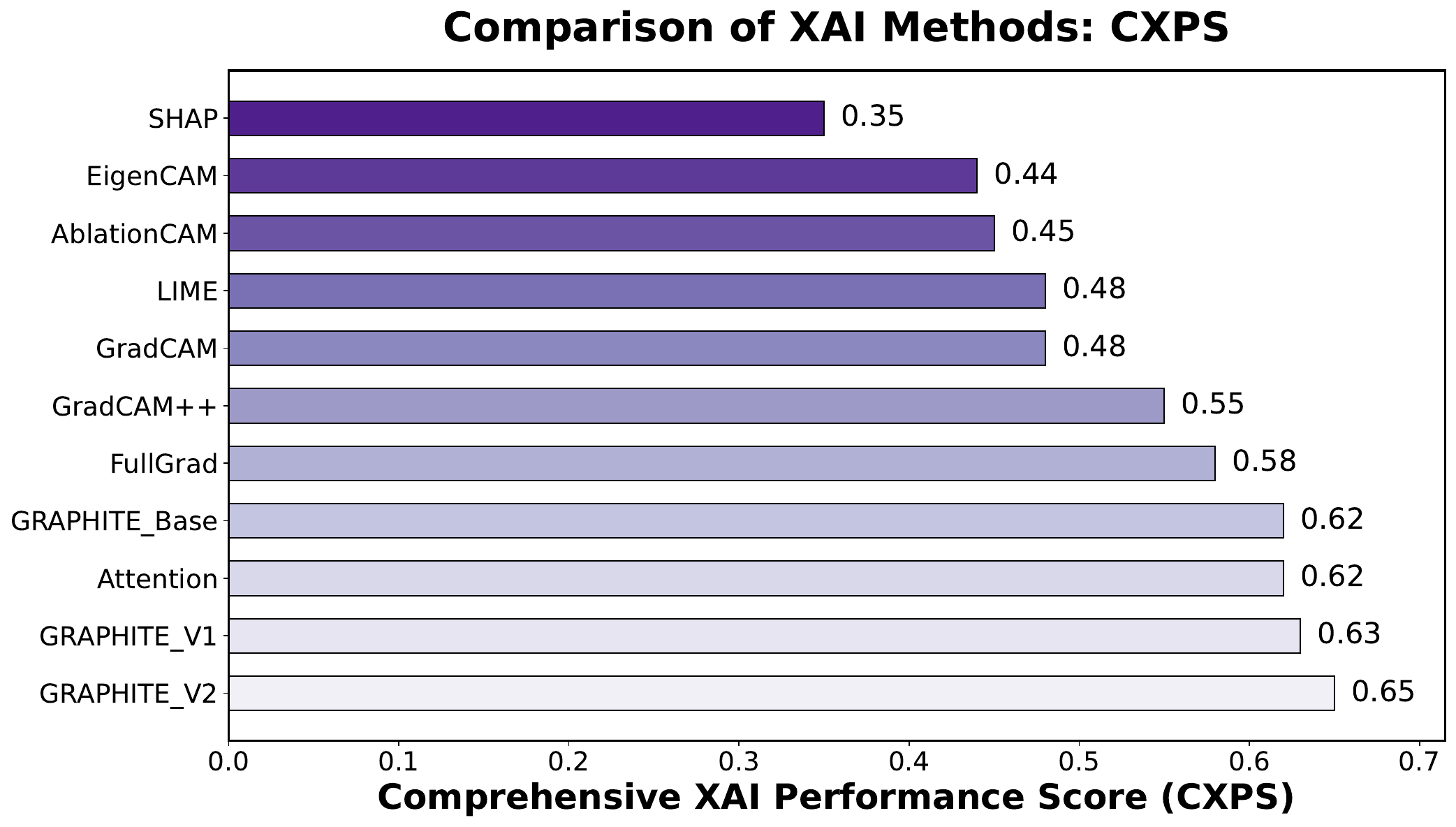}
\caption{Comparison of XAI methods based on the CXPS metric. The GRAPHITE variants (GRAPHITE-Base, GRAPHITE-V1, GRAPHITE-V2) achieve the highest CXPS scores, with GRAPHITE-V2 scoring 0.65, indicating its superior balance of interpretability and predictive performance. Other methods, such as Attention and FullGrad, perform moderately well, while EigenCAM, AblationCAM and GradCAM achieve comparatively lower CXPS scores. Model-agnostic methods LIME (CXPS=0.48) and particularly SHAP (CXPS=0.35) score lowest, reflecting limitations in overall performance and interpretability coherence in this application according to our composite metric. This evaluation highlights GRAPHITE's advantage in delivering a reliable and interpretable solution for breast cancer TMA analysis.}
\label{fig:cxps}
\end{figure}

\begin{figure}[htbp]
\includegraphics[width=0.9\textwidth]{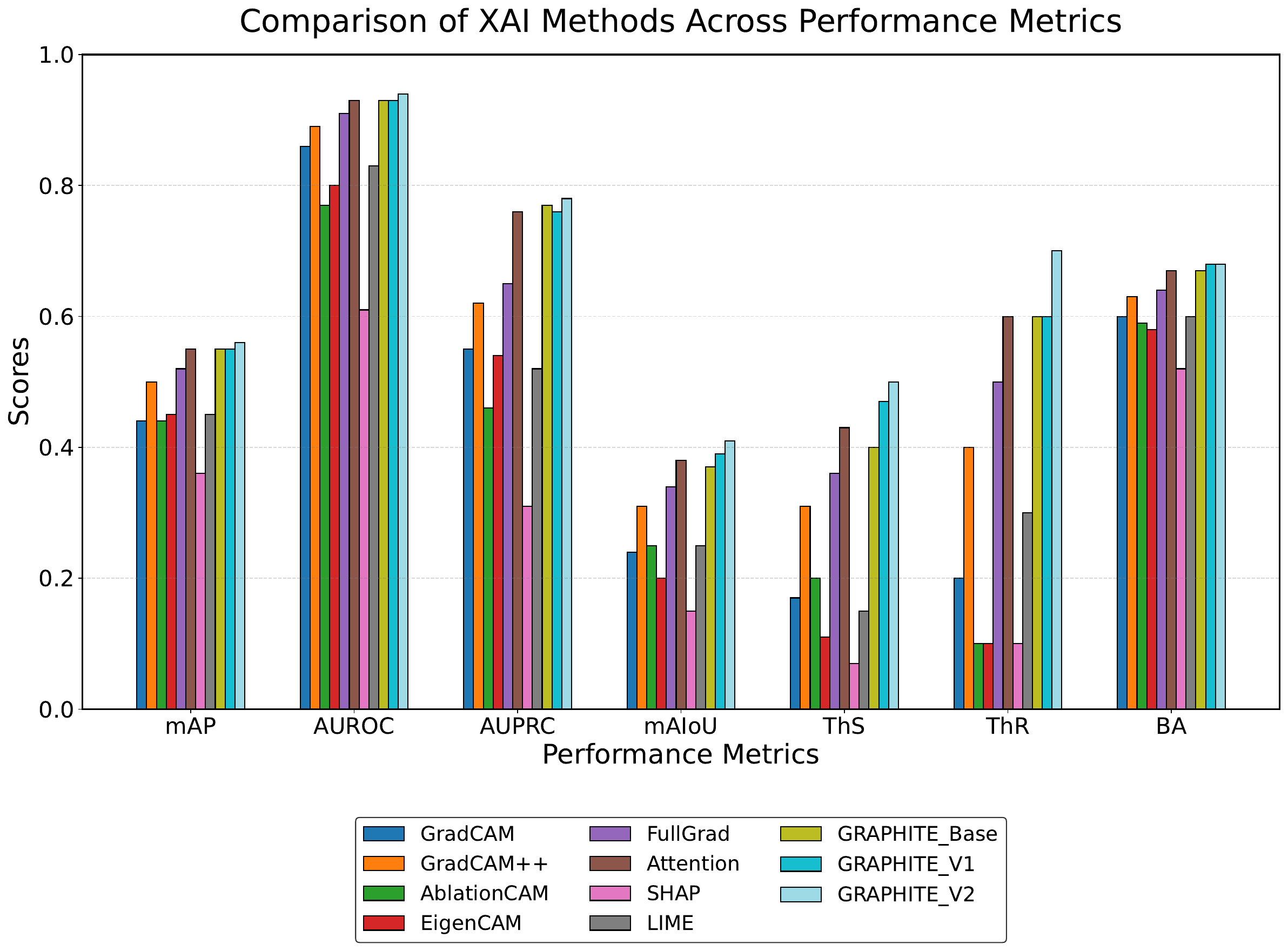}
\caption{Comparison of various XAI methods across multiple performance metrics. The GRAPHITE variants GRAPHITE-V2, GRAPHITE-V1 and GRAPHITE-Base demonstrate superior performance across most metrics, particularly AUROC and AUPRC, indicating their effectiveness in both discriminative power and precision-recall balance. GRAPHITE-V2 achieves the highest scores in AUROC, AUPRC and mIoU, highlighting its strength in interpretability and robustness. Other methods such as FullGrad and Attention, perform well but consistently fall short of the GRAPHITE models, especially in mIoU and robustness metrics (ThS and ThR). This comprehensive evaluation demonstrates the advantages of the GRAPHITE framework in delivering both accurate and interpretable visualisations, and suitabilityfor clinical decision-making in breast cancer tissue analysis. mAP: mean average precision. AUROC: area under the receiver operating characteristic curve. AUPRC: area under the precision-recall curve. mIoU: mean intersection over union. ThS: threshold stability. ThR:  threshold robustness. BA: balanced accuracy.}
\label{bargraph}
\end{figure}

\subsubsection{Net Benefit Curve (NBC)}
The NBC evaluates the clinical utility of predictive models across a range of decision thresholds. It is particularly relevant in clinical decision-making scenarios, where the cost of FPs and FNs varies depending on the threshold chosen. The net benefit (NB) at a specific threshold represents a balance between TP predictions and the penalty for FP predictions, adjusted for the chosen threshold \cite{10.3390/healthcare11162244}. The formula for calculating NB is:
\begin{equation}
    \text{Net Benefit} = \frac{\text{TP}}{\text{Total}} - \frac{\text{FP}}{\text{Total}} \cdot \frac{\text{Threshold}}{1 - \text{Threshold}}
\end{equation}
where $\text{Total}$ is the total number of patients. NB at a given threshold quantifies the proportion of TPs adjusted by the harm of FPs relative to the threshold. This helps assess whether the model adds value compared to ``no net benefit''. 

The NBC plots NB over a range of thresholds, allowing clinicians and researchers to visually compare the utility of different models in a practical setting. The area under the NBC (AUDC) provides a single quantitative measure of model utility across thresholds. Higher AUDC values indicate that the model maintains a beneficial balance of TPs and penalises FPs effectively across a wider threshold range, thus demonstrating superior clinical utility. In this study, the NBC was used to compare the effectiveness of different XAI methods, guiding the selection of models that optimise patient outcomes at various decision thresholds.

\subsection{Base Classification Performance}
Prior to evaluating the explainability aspects, we assessed the performance of the base classification model in distinguishing between tumour and normal tissue in TMA cores. The Stage 1 MIL-based classification model demonstrates excellent discriminative capability, achieving an AUROC of 0.96. The model also exhibits strong precision-recall performance with an F1 score of 0.98. These results indicate the model's robust ability to differentiate between tumour and normal tissue patterns across diverse TMA samples, establishing a reliable foundation for subsequent explainability analysis, ensuring that GRAPHITE's interpretations are based on accurate diagnostic predictions.

\subsection{Comparison of GRAPHITE Variants}
Next we evaluated three variants of GRAPHITE designed with increasing levels of interpretability and robustness features. The first, GRAPHITE-Base, employs a multilevel fusion approach to integrate information across different magnification levels in tissue images, establishing a foundational level of interpretability. Building on this, GRAPHITE-V1 introduces MIL attention, allowing the model to highlight critical areas within each magnification level, which aligns more closely with diagnostic reasoning by focussing on regions of interest. The most advanced variant, GRAPHITE-V2, incorporates both MIL attention and FullGrad, a gradient-based technique that further enhances interpretability by refining the focus on diagnostically relevant tissue regions. The addition of FullGrad in GRAPHITE-V2 provides improved localisation and stability, making it the most robust and interpretable variant within the GRAPHITE framework.

Of these variants, GRAPHITE-V2 indeed achieved the highest performance across all metrics (see the last three rows of Table~\ref{tab:performance-comparison}). Specifically, the mIoU score of 0.41 indicates that GRAPHITE-V2 accurately localises regions of interest, aligning well with pathologists’ diagnostic areas. Additionally, GRAPHITE-V2 shows strong robustness across decision thresholds, achieving the highest scores in ThS of 0.50 and ThR of 0.70, underscoring its stability in clinical applications where decision thresholds may vary. The superiority of GRAPHITE-V2 is further confirmed by the highest mAP of 0.56, AUROC of 0.94 (see also Figure~\ref{auroc}), AUPRC of 0.78 (see also Figure~\ref{auprc}), and BA of 0.68, although in terms of the latter metrics it performed quite similar to the other variants. Finally, the comprehensive CXPS score of 0.65 for GRAPHITE-V2 highlights its balanced performance across both interpretability and predictive metrics, making it the most effective model among the evaluated XAI methods (Figure~\ref{fig:cxps}).


Comparing GRAPHITE-V2 to GRAPHITE-V1 and GRAPHITE-Base, we observe that the addition of FullGrad in GRAPHITE-V2 leads to noticeable improvements in interpretability and robustness metrics. GRAPHITE-V1, which combines multilevel fusion with MIL attention but lacks FullGrad, achieves an AUROC of 0.93 and an AUPRC of 0.76, slightly lower than GRAPHITE-V2 but still outperforming most traditional methods. The mIoU score of 0.39 for GRAPHITE-V1 indicates strong region localisation, though it is slightly less precise than GRAPHITE-V2. GRAPHITE-Base, which employs only multilevel fusion, achieves competitive performance with an AUROC of 0.93 and AUPRC of 0.77. However, the absence of MIL attention and FullGrad in GRAPHITE-Base results in a lower ThS of 0.40 and ThR of 0.60, indicating reduced stability and robustness across decision thresholds. These findings demonstrate the incremental benefits of integrating MIL attention and FullGrad into the GRAPHITE framework.
\begin{figure}[!t]
\centering
\includegraphics[width=\textwidth]{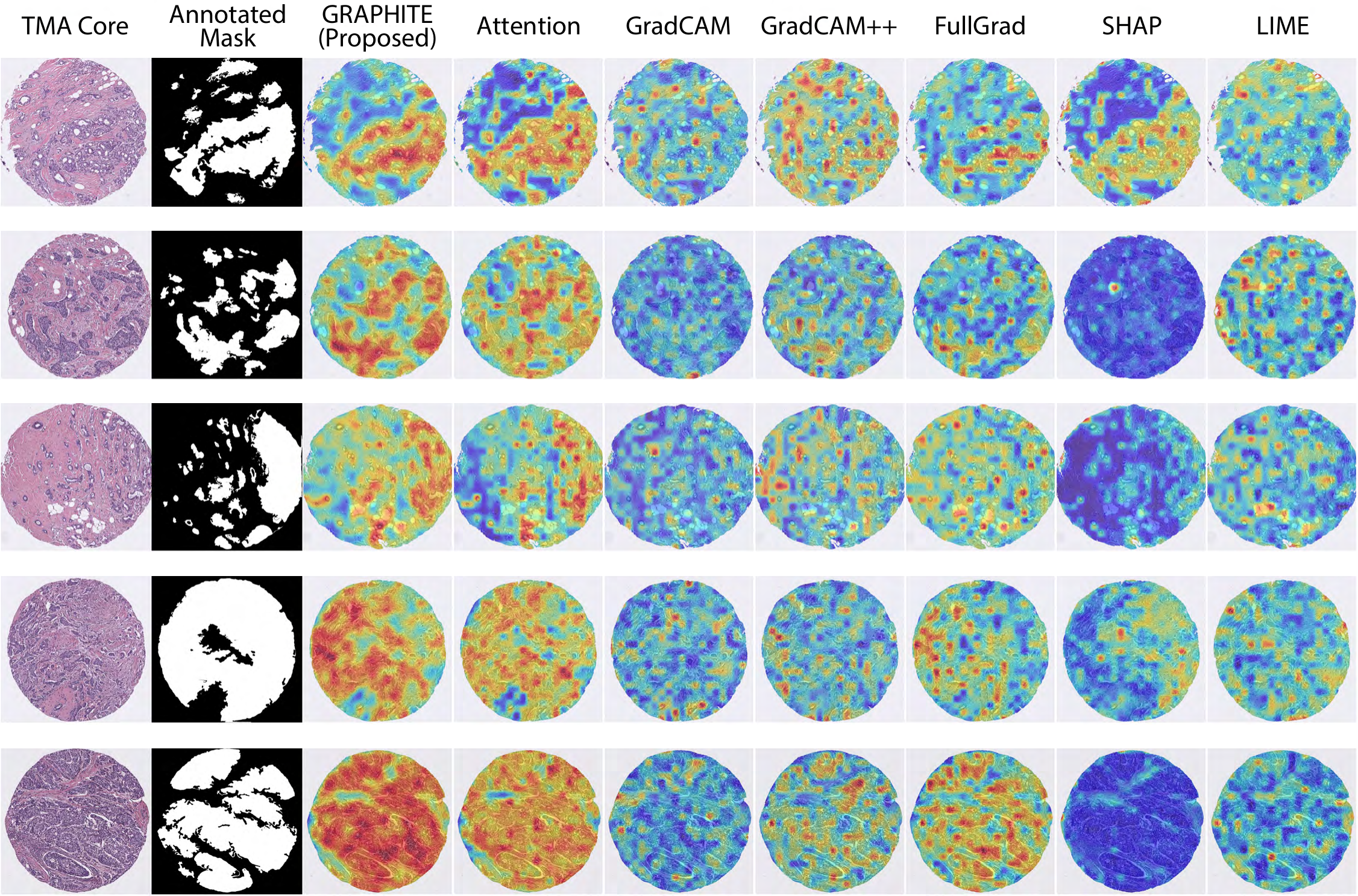}
\caption{Part 1 of 2. Comparative visualization of breast cancer TMA cores using GRAPHITE, attention-based methods, gradient-based methods (GradCAM, GradCAM++, FullGrad), and model-agnostic methods (LIME and SHAP), alongside the original TMA core images and pathologist-annotated masks. Each row corresponds to a different TMA core. GRAPHITE demonstrates improved alignment with the annotated masks, enhancing interpretability by better capturing relevant tumour regions. In contrast, SHAP visualizations appear more diffuse and less focused on primary tumour areas.}
\label{fig:heatmap_comparison_1}
\end{figure}

\subsection{Comparison With Traditional XAI Methods}
To comprehensively evaluate GRAPHITE's performance, we compared it against several state-of-the-art XAI methods (See Table~\ref{tab:performance-comparison} and Figure~\ref{bargraph}), ensuring a robust and fair comparison.

Gradient-based methods, such as GradCAM and its advanced variant, Grad-CAM++, which utilize class-specific gradient information, achieved moderate AUROCs of 0.86 and 0.89 respectively, demonstrating their foundational class-discriminative capabilities. FullGrad, which extends gradient-based approaches by incorporating both input and neuron sensitivity components, showed improved performance (AUROC: 0.91) but exhibited limitations in localization precision metrics (mAP: 0.52, mIoU: 0.34). EigenCAM's principal component analysis approach and AblationCAM's feature removal strategy demonstrated comparatively lower performance across all metrics, suggesting potential limitations in their underlying theoretical frameworks for this complex task.

Attention-based methods, like MIL-Attention (representing the attention mechanism within the classifier itself), establish a competitive baseline (AUROC: 0.92, AUPRC: 0.77, ThR: 0.60), matching GRAPHITE-Base's performance. Model-agnostic methods LIME and SHAP were also evaluated. LIME achieved moderate results (AUROC: 0.83, AUPRC: 0.52), outperforming basic GradCAM but falling short of more advanced methods. SHAP, in contrast, demonstrated limited performance across most metrics (AUROC: 0.61, AUPRC: 0.31, mIoU: 0.15). This suboptimal performance may be attributed to inherent limitations in SHAP's DeepExplainer, which is built upon the DeepLIFT algorithm. Although, DeepLIFT introduces rules (such as the RevealCancel rule and difference-based propagation) to manage non-linearities and saturation, its assumptions may not hold in the presence of complex network behaviours\cite{DEEPLIFT}. This may lead to oversimplified explanations that fail to capture spatial dependencies and non-linear patterns intrinsic to histopathological images. These results suggest that while model-agnostic methods offer flexibility, they may not capture the nuanced, spatially correlated features critical for histopathology interpretation.

\begin{figure}[!t]
\centering
\includegraphics[width=\textwidth]{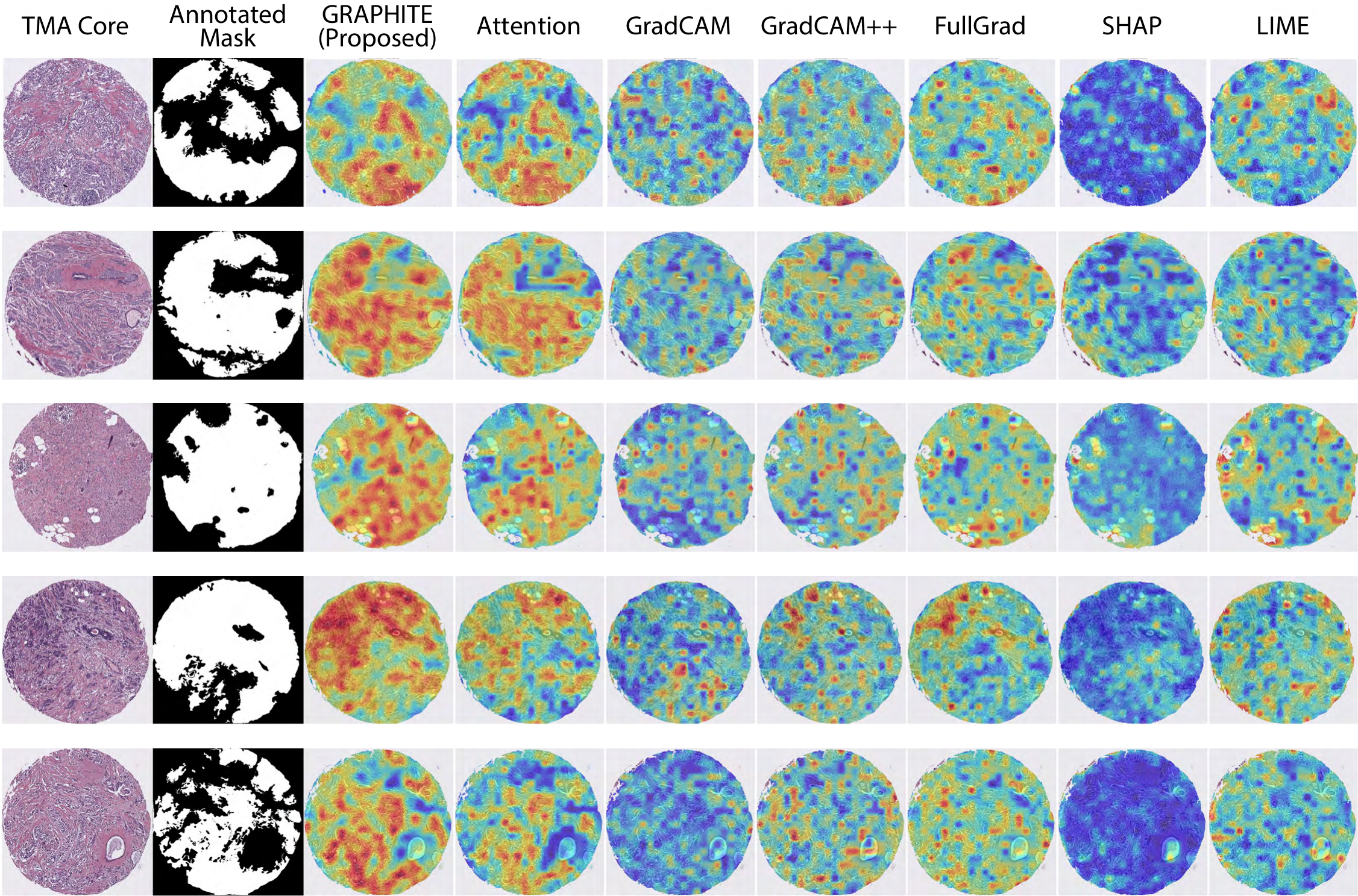}
\caption{Part 2 of 2. Continuation of comparative visualizations from Figure~\ref{fig:heatmap_comparison_1}, showing additional TMA core examples analyzed using the same set of interpretation techniques. Observations are consistent with Figure~\ref{fig:heatmap_comparison_1}: GRAPHITE continues to show better alignment with pathologist annotations, while SHAP remains less localized in identifying tumour regions.}
\label{fig:heatmap_comparison_2}
\end{figure}

GRAPHITE-V1 advances the discriminative capability over the base attention with an AUROC of 0.93, though with a marginally lower AUPRC (0.76). GRAPHITE-V2 ultimately achieves superior performance across all metrics (AUROC: 0.94, AUPRC: 0.78, ThS: 0.50, ThR: 0.70), highlighting the effectiveness of its integrated architectural innovations (multiscale graph structure, GAT, SAN, and confidence-based fusion) in delivering both reliable predictions and interpretable visualizations  (Figures~\ref{fig:heatmap_comparison_1} and~\ref{fig:heatmap_comparison_2}).

\begin{figure}[!t]
\centering
\includegraphics[width=0.8\textwidth]{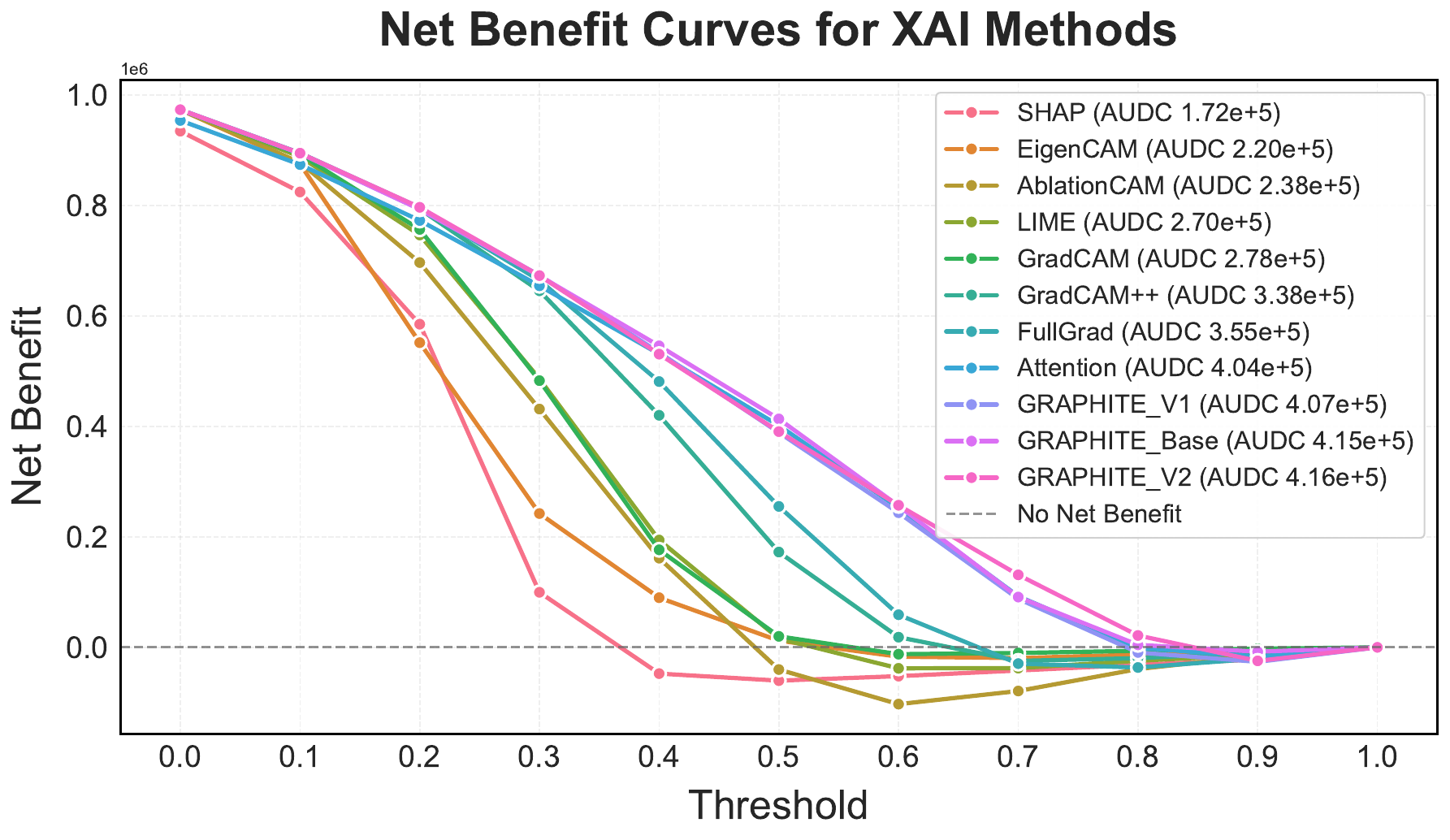}
\caption{Net Benefit Curves for various XAI methods, illustrating the clinical utility of each method across a range of decision thresholds. The net benefit is plotted as a function of threshold, showing the trade-off between true positives and the cost of false positives for each method. The AUDC is provided in the legend for each method as a quantitative measure of the model's utility. GRAPHITE-V2 achieves the highest AUDC (4.16e+5), followed by GRAPHITE-V1 (4.08e+5) and GRAPHITE-Base (4.15e+5), indicating superior net benefit across thresholds. These results highlight the advantage of the GRAPHITE variants over other XAI methods, such as Attention (AUDC = 4.04e+5) and FullGrad (AUDC = 3.55e+5).SHAP (AUDC = 1.72e+5) show considerably lower clinical utility according to this analysis. The dashed line represents the ``no net benefit'' baseline, where a model yields no additional benefit over treating all or none of the cases. The higher net benefit values for GRAPHITE variants underscore their potential effectiveness in clinical decision-making for breast cancer analysis.}
\label{fig:netbenefit}
\end{figure}

\subsection{Clinical Utility and Net Benefit Analysis}
The clinical utility of the evaluated XAI methods was further assessed using NBC, with the AUDC metric serving as a quantitative measure of clinical effectiveness. GRAPHITE-V2 achieved the highest AUDC of 4.17e+5, indicating the highest net benefit across decision thresholds (Figure \ref{fig:netbenefit}). This result highlights the potential of GRAPHITE-V2 to improve clinical decision-making by balancing true positive predictions with a minimised cost of false positives. GRAPHITE-V1 and GRAPHITE-Base also show high AUDC values, further validating the robustness of the GRAPHITE framework. Traditional methods such as GradCAM and FullGrad demonstrated lower AUDC scores, while the model-agnostic methods showed considerably less utility in this clinical context. LIME achieved an AUDC of 2.70e+5, and SHAP performed poorest with an AUDC of 1.72e+5. These low values for LIME and especially SHAP align with their performance in other metrics and underscore their limited effectiveness in guiding threshold-dependent clinical decisions for this specific task compared to the GRAPHITE variants.

In summary, the results indicate that GRAPHITE-V2 is the most effective XAI method among those evaluated, achieving high scores across interpretability, predictive accuracy and robustness metrics. The integration of multilevel fusion, MIL attention and FullGrad enables GRAPHITE-V2 to offer fine-grained and clinically meaningful visual explanations, setting it apart from other methods in both performance and clinical applicability. The comprehensive evaluation demonstrates the potential of GRAPHITE to provide interpretable and accurate AI-driven insights for breast cancer analysis, making it a valuable tool in computational pathology.

\section{Discussion}
GRAPHITE, a novel post hoc explainability framework proposed in this paper, addresses critical challenges in interpretable AI for digital pathology. Our results demonstrate significant improvements in both model performance and interpretability compared to existing XAI methods, with important implications for clinical practice and future developments in computational pathology.

GRAPHITE's superior performance across multiple evaluation metrics, particularly its achievement of 0.94 AUROC and 0.78 AUPRC, represents a substantial advancement in interpretable AI for histopathology analysis. The framework's innovative integration of multiscale hierarchical graph representation with confidence-based adaptive fusion mechanisms has proven particularly effective in capturing both fine-grained cellular detail and broader tissue-level patterns. This comprehensive approach addresses a fundamental limitation of existing methods, which often struggle to balance local and global feature interpretation in complex histology images.

From a clinical perspective, GRAPHITE's robust performance has significant implications for pathology workflows. The framework's ability to provide detailed, multiscale visualisations aligns well with pathologists' diagnostic processes, potentially reducing the time required for analysis while maintaining high accuracy. The superior threshold stability (ThS = 0.50) and robustness (ThR = 0.70) of GRAPHITE-V2 suggest that the model can maintain consistent performance across varying clinical settings and operator preferences, addressing a critical requirement for clinical deployment. Furthermore, the high mean intersection over union (mIoU = 0.41) indicates strong alignment with expert annotations, suggesting potential utility in reducing interobserver variability among pathologists.

The clinical utility of GRAPHITE is further validated through decision curve analysis, which demonstrates the framework's robust performance across varying diagnostic thresholds. This reveals GRAPHITE's superior ability to balance the benefits of true positive predictions against the costs of false positives---a critical consideration in clinical pathology where diagnostic decisions often require different sensitivity-specificity trade-offs. The sustained performance advantage of all GRAPHITE variants at higher thresholds is particularly noteworthy, as it indicates reliability in scenarios where high specificity is crucial for avoiding unnecessary interventions. This is especially valuable in breast cancer diagnosis, where false positives can lead to unnecessary procedures and patient anxiety.

When compared to existing XAI methods, GRAPHITE's advantages become particularly apparent. Traditional approaches like GradCAM (AUROC = 0.86) and attention-based methods (AUROC = 0.92) can struggle to provide consistent interpretations across different scales and tissue types. Similarly, widely used model-agnostic methods such as LIME and SHAP—though flexible—demonstrated limitations in our evaluation (e.g., SHAP AUROC = 0.61), likely due to challenges in capturing complex spatial dependencies within intricate tissue patterns. GRAPHITE's hierarchical approach and adaptive fusion mechanism appear better suited to address these histopathology-specific challenges, resulting in more reliable and clinically meaningful visualisations. The framework's superior performance across discriminative power, interpretability metrics, and clinical utility assessments (AUDC) compared to this broad range of techniques suggests that it successfully bridges the gap between model accuracy and clinical utility, a crucial consideration for real-world applications.

The broader implications of GRAPHITE for XAI in healthcare are significant. Its success in providing interpretable, multiscale analysis demonstrates the potential for AI systems to augment rather than replace clinical expertise. The high CXPS score (0.65) suggests that GRAPHITE effectively balances technical performance with practical utility, a crucial consideration in building trust in AI-assisted diagnostic systems. This balance is particularly important in pathology, where interpretability is essential for clinical adoption and regulatory compliance.

Despite these advances, the proposed framework has several limitations. First, the computational complexity of the hierarchical graph construction and multiscale analysis may present challenges for real-time applications. Future optimisations could focus on reducing this computational overhead while maintaining performance. Second, while GRAPHITE shows excellent performance on breast cancer TMAs, its generalisability to other cancer types and whole slide tissue images requires further investigation. Additionally, the framework could benefit from incorporating temporal consistency in whole-slide image analysis and extension to multiclass classification scenarios.

The mentioned limitations suggest several promising avenues for future research. The proposed architecture could be extended to incorporate additional modalities, such as immunohistochemistry or molecular data, potentially enabling more comprehensive diagnostic insights. Furthermore, the principles underlying GRAPHITE's multiscale analysis and confidence-based fusion could be adapted to other medical imaging domains, where interpretability and robust performance are equally crucial.

The success of GRAPHITE in breast cancer TMA analysis represents a significant step toward more interpretable and clinically useful AI systems in pathology. By effectively addressing the challenges of multiscale feature integration and providing robust interpretable visualisations, GRAPHITE demonstrates the potential for AI to enhance rather than complicate clinical decision-making. As digital pathology continues to evolve, frameworks like GRAPHITE will play an increasingly important role in ensuring that AI systems remain both powerful and interpretable, ultimately contributing to improved patient care.

\section{Conclusion}
The GRAPHITE framework presented in this study addresses critical challenges in the interpretability and clinical applicability of AI models in the analysis of breast cancer tissue microarrays. By integrating multiscale hierarchical graph representation, MIL attention and FullGrad, GRAPHITE achieves high accuracy and interpretable visualisations that align with the pathologists' diagnostic reasoning. Our results indicate that GRAPHITE, particularly its most advanced variant (GRAPHITE-V2), outperforms traditional XAI methods in both predictive performance and robustness. GRAPHITE shows promise as a reliable tool for enhancing clinical decision support in computational pathology, supporting more precise and interpretable diagnostics in oncology.

\section*{Data Availability}
The TMA image dataset and benign whole slide images (WSI) are not publicly available due to ethics restrictions; however, they may be accessed upon reasonable request to E.K.A.M.

\section*{Code Availability}
Our work is fully reproducible and source code is available at \href{https://github.com/raktim-mondol/GRAPHITE}{Github}.

\section*{Institutional Review Board Statement} 
Ethical approval for this study was provided by the South Eastern Sydney Local Health District (SESLHD) Human Research Ethics Committee (HREC) at Prince of Wales Hospital, Sydney, Australia. Approval for the TMA cores from the randomised radiotherapy clinical trial was granted under HREC 96/16, and approval for the benign breast WSIs was obtained under HREC/17/POWH/389-17/176. All patients recruited to the trial provided informed consent. All methods were performed in accordance with relevant institutional guidelines and regulations.

\section*{Informed Consent Statement}
Informed consent was obtained from all subjects involved in the study.

\section*{Acknowledgement}
This research was undertaken with the assistance of resources and services from the National Computational Infrastructure (NCI), which is supported by the Australian Government. Additionally, data preprocessing was performed using the computational cluster Katana, which is supported by Research Technology Services at UNSW Sydney.

\footnotesize
\bibliographystyle{ieee_tr_custom.bst}  
\bibliography{references}  

\end{document}